\documentclass[apj,twocolumn]{emulateapj}
\usepackage{amsmath}
\usepackage{txfonts}
\usepackage{amssymb}
\usepackage{graphicx}
\usepackage{color}
\definecolor{orange}{rgb}{1.0,0.4,0.0}
\definecolor{purple}{rgb}{0.5,0.0,0.5}
\definecolor{gray}{rgb}{0.7,0.75,0.71}
\usepackage{longtable}
\usepackage{rotating}
     \usepackage[unicode,citecolor=blue]{hyperref}
\usepackage[all]{hypcap}
\usepackage{booktabs}
\usepackage{natbib}
\usepackage{threeparttable}
\usepackage[10pt]{moresize}

\newcommand{\msolar}{\,$\rm M_{\odot}$} 
\newcommand{\lsolar}{\,$\rm L_{\odot}$}

\newcommand{\kms}{km s$^{-1}$}

\renewcommand\email\texttt
\def\spose#1{\hbox to 0pt{#1\hss}}
\def\lta{\mathrel{\spose{\lower 3pt\hbox{$\sim$}}
    \raise 2.0pt\hbox{$<$}}}
\def\gta{\mathrel{\spose{\lower 3pt\hbox{$\sim$}}
    \raise 2.0pt\hbox{$>$}}}

\begin{document} 

\shorttitle{The orbital nature of ellipsoidal red giant binaries}
\shortauthors{J. D. Nie et al.}

\title{The Orbital Nature of 81 Ellipsoidal Red Giant Binaries in the 
Large Magellanic Cloud}
\author{
J. D. Nie\altaffilmark{1},
P. R. Wood\altaffilmark{2},
C. P. Nicholls\altaffilmark{3}}
\altaffiltext{1}{Key Laboratory of Optical Astronomy, National Astronomical Observatories, 
Chinese Academy of Sciences, Beijing 100012, China; \email{jdnie@bao.ac.cn}}
\altaffiltext{2}{Research School of Astronomy and Astrophysics, Australian National
University, Canberra, ACT 2611, Australia; \email{peter.wood@anu.edu.au}}
\altaffiltext{3}{Institute for Astrophysics, University of Vienna, T\"{u}rkenschanzstra{\ss}e 17, 1180 Wien, Austria;
\email{christine.nicholls@univie.ac.at}}

\begin{abstract}\label{abstract}
In this paper, we collect a sample of 81 ellipsoidal red giant 
binaries in the LMC and we study their orbital natures individually and 
statistically.  The sample
contains 59 systems with circular orbits and 22 systems with eccentric
orbits. We derive orbital solutions using the 2010 version of the
Wilson--Devinney code. The sample is selection-bias
corrected and the orbital parameter distributions are compared
to model predictions for the LMC and to observations in the solar
vicinity.  The masses of the red giant primaries are found to range from
about 0.6--9\msolar~with a peak at around 1.5\msolar, in agreement with 
studies of the star formation history of the LMC which
find a burst of star formation beginning around 4\,Gyr ago.
The observed distribution of mass ratios $q = m_{\rm 2}/m_{\rm 1}$ 
is more consistent with the flat $q$ distribution derived for the 
solar vicinity by \citet{2010ApJS..190....1R}
than it is with the solar vicinity $q$ distribution derived by \citet{1991A&A...248..485D}.  There is no
evidence for an excess number of systems with equal mass components. 
We find that about 20\% of the ellipsoidal binaries have eccentric orbits, twice the
fraction estimated by \citet{2004AcA....54..347S}.  Our eccentricity
evolution test shows that the
existence of eccentric ellipsoidal red giant binaries on the upper 
parts of the red giant branch (RGB) can only be explained
if tidal circularization rates are $\sim$1/100 the rates
given by the usual theory of tidal dissipation in convective stars.

\end{abstract}


\section[]{Introduction}

Ellipsoidal red giant binaries are close binary systems consisting of
a distorted red giant and, in most cases, an unseen companion. The red
giant is distorted into a roughly ellipsoidal shape 
because it fills a large fraction of its Roche lobe.  Orbital rotation 
of the distorted red giant gives rise to a characteristic change in apparent 
brightness, with two light maxima and two light minima in each orbital 
period in contrast to the radial velocity variation which has one maximum 
and one minimum per orbit 
\citep[e.g.][]{1985ApJ...295..143M,2010MNRAS.405.1770N,2014AJ....148..118N}.

The ellipsoidal red giant binaries in the Large Magellanic Cloud (LMC)
are also known as sequence E stars because of another characteristic
property -- these stars lie on a distinct linear Period--Luminosity
(PL) sequence named sequence E \citep{1999IAUS..191..151W,
  2004AcA....54..347S}.  The position of this sequence in the PL plane
can vary between different publications depending on whether the period
adopted is the orbital period or the period between light maxima
\citep[e.g.][]{2000PASA...17...18W,
  2004MNRAS.353..705I,2007AcA....57..201S,2008AJ....136.1242F}.  
The orbital periods of sequence E stars from the
above papers lie in the range $\sim$ 40--1000 days
while apparent $K$ magnitudes lie in the range $\sim$12--16
although fainter and shorter-period ellipsoidal
variables do exist on the red giant branch in the LMC 
\citep{2014AcA....64..293P}.

The PL sequence is basically a sequence of binary systems born with
different separations.  A red giant almost fills its Roche
lobe in an ellipsoidal binary system and the orbital separation
of the two stars is typically 3-4 times the radius of the red
giant.   A larger orbital separation thus requires a larger radius $R$ and hence a
higher luminosity $L$ for the red giant.   Since a larger separation also 
leads to a larger orbital period via  
Kepler's third law, a PL sequence for ellipsoidal red giant binaries results
\citep[e.g. see Figure 4 of][where
the observed period used for the sequence E stars is the time
between light maxima (the photometric period), which is half the orbital
period]{1999IAUS..191..151W}.  Mass variations of the components of
the binary and variations in the Roche lobe filling factor both
lead to a scatter in the PL relation \citep{2004AcA....54..347S}.

In ellipsoidal red giant binaries, the red giant is an AGB or RGB star
undergoing radius expansion as it evolves and the red giant is about to 
fill its Roche lobe. At the same time, the unseen companion, which is 
usually a main sequence star of lower mass, keeps its size and luminosity
relatively small and unchanged. When the red giant fills its Roche
lobe, the companion gains mass from the donor red giant, usually in a
run-away process, and this kind of dynamical mass transfer builds up a
common envelope around the two bodies.  The common envelope event
quickly creates a close binary planetary nebula (PN), a binary
post--AGB star, a binary post-RGB star or a merged star that will
continue to evolve as a red giant \citep{2012MNRAS.423.2764N}.
An alternative stable mass transfer process on an evolutionary timescale
can occur for a small range in the mass ratio of the two stars in 
the binary system \citep[e.g.][]{2002ASPC..279..297H}.

In order to make population formation studies of the descendants
of the interacting binary systems mentioned above, 
a knowledge of the orbital properties and frequency of occurrence
of their parent systems is essential. Since the ellipsoidal red giant
binaries are the immediate precursor binaries of above descendants, a
knowledge of their orbital properties in the population under study
should put strict constraints on binary population simulation models.
In particular, we need to know properties such as the statistical
distributions of the binary masses $m_1$ and $m_2$, the mass ratio
$q=m_2/m_1$, the orbital separation $a$ and the eccentricity $e$.  
All these properties can be obtained for ellipsoidal red giant binaries
provided light and velocity curves exist.

The LMC is an ideal stellar population to study because, in addition
to a well-determined population of ellipsoidal red giant binaries
(from the MACHO and OGLE II microlensing experiments),
there exist quite well known populations of many other stellar types
such as RGB stars, AGB stars, post-RGB stars, post-AGB stars and
planetary nebulae.  Thus the populations predicted by binary
population modelling can be compared to the observed population.  Such
modelling was done by \citet{2012MNRAS.423.2764N} but they adopted
orbital element distributions from the solar vicinity studies of
\citet{1991A&A...248..485D} and \citet{2010ApJS..190....1R}.  It is
desirable to see if the same distributions exist in samples of
binaries in the LMC where there is a different metallicity
distribution and a different star formation history.  The LMC
ellipsoidal red giant variable population therefore provides an opportunity to
directly determine orbital element distributions in LMC binary
systems.  This is the aim of this paper.

To determine the complete orbital properties of ellipsoidal red giant
binaries at a known distance (which is the case for LMC binaries), 
light curves and radial velocity data are sufficient
(see \autoref{best_sol} of this paper for a justification). 
Examples of the derivation of complete orbital parameters
for ellipsoidal variables at a known distance are given in
\citet{2009ApJ...702..403W} and \citet{2012MNRAS.421.2616N}.  
Thanks to the MACHO and OGLE
projects, extensive light curve data are available for LMC ellipsoidal
red giant binaries, along with good estimates of the populations of
these systems. However, radial velocity data are rare because it is
very time consuming to cover even one orbit given the long orbital
periods and only a few radial velocity studies have been carried out up 
to now. \citet{2010MNRAS.405.1770N} took radial velocity observations
of 11 ellipsoidal variables in the LMC and roughly estimated orbital
parameters for the systems. However, for a complete understanding of those
objects, a more accurate parameter determination is needed. As a
special study of the ellipsoidal binaries, \citet{2012MNRAS.421.2616N}
monitored radial velocities for 7 highly eccentric systems in the LMC
and determined their orbital parameters with the 2010 version of the
Wilson--Devinney (hereinafter WD) code \citep{1971ApJ...166..605W,
  1979ApJ...234.1054W, 1990ApJ...356..613W,2009ApJ...702..403W}, which
can produce complete orbital solutions if the absolute luminosity is
known, as it is for stars in the LMC.  That paper is the only existing work that
gives complete solutions for LMC ellipsoidal variables.

In order to have larger samples, and to select objects with as wide a
set of parameters as possible, \citet{2014AJ....148..118N} carried out
radial velocity observations of 80 ellipsoidal red giant binaries in
the LMC.  They provided radial velocity curves for individual stars
as well as some statistics on the observed stellar parameters.
However, orbital parameters were not determined for these binaries and
this is the main goal of this paper.  We use the 2010 version of the
WD code to attempt modelling of 79 ellipsoidal variables in 
\citet{2014AJ....148..118N}.  In addition, we attempt modelling of the 
11 objects in \citet{2010MNRAS.405.1770N} in order to better determine 
their orbital parameters. Together with the 7 eccentric systems in 
\citet{2012MNRAS.421.2616N} 
whose orbital solutions we also attempt to re-derive using the methods in
this paper for consistency, we have 
an initial sample of 97 ellipsoidal binaries. 
(We only modelled 79 of the 80 objects in \citet{2014AJ....148..118N} 
as one object had already been observed by \citet{2012MNRAS.421.2616N}.) 
The complete sample can provide us with statistical information about the 
properties of LMC binaries with intermediate initial separations such that 
binary interactions occur on the RGB or AGB.  These properties can be
compared to the properties of similar binaries in the solar vicinity.
In addition, by comparing the observed binary properties with binary
evolution models, we can investigate the assumptions of the evolution
models such as the effect of tides on eccentricity evolution.
 
\section[]{Observational data}

In order to model the orbits of the LMC ellipsoidal variables, we
require both light and velocity curves.  The light curve data we use
are mainly from OGLE II
\citep{2005AcA....55...43S,1997AcA....47..319U,2004AcA....54..347S}, 
sometimes supplemented by OGLE III data if it is published.  The
light curves are in the $I$ band.  The radial 
velocities are provided by \citet{2014AJ....148..118N} for 79 ellipsoidal
variables, by \citet{2010MNRAS.405.1770N} for their 11 ellipsoidal
variables and by \citet{2012MNRAS.421.2616N} for their 7 eccentric binaries.
The light curve photometry, supplemented by $K$-band photometry
from the Two Micron All Sky Survey (2MASS) catalog
\citep{2003yCat.2246....0C}, provides the $K$ magnitude and the $I-K$ 
color.  From these and the adopted LMC distance modulus of 
18.49 \citep{2014AJ....147..122D} and reddening E($B$-$V$)$ = 0.08$ 
\citep{2006ApJ...642..834K}, the basic stellar parameters  
effective temperature $T_{\rm eff1}$, luminosity $L_{1}$, and the mean 
radius $R_{1}$ of the primary star can be computed (see 
\citealt{2014AJ....148..118N} for the details of the calculation of these 
quantities).  The light and velocity curve data also allow derivation 
of the orbital period $P$ and the radial velocity semi-amplitude $K_1$ 
of the red giant.  

\section[]{Modelling details}\label{method}

In order to get a converged solution in the WD code, a reasonably
accurate approximation to the final solution is required as input.
This is especially so when a large number of solutions is sought as is
the case here.  Hence, we used a separate code to estimate the initial 
orbital parameters.  This code uses the equations below to determine
an approximate solution that is fed into the WD code for iterative
refinement.  The eccentricity is assumed to be zero. 
Firstly, we have Kepler's Law
\begin{equation}\label{kep_eqn}
\frac{m_{1}(1+q)}{a^3}=\frac{4 \pi^2}{G P^2}
\end{equation}
where $m_1$ is the mass of the red giant, $q$ is the mass ratio $m_2/m_1$, $m_2$ 
is the mass of the unseen companion, $a$ is the semi-major axis of the orbit, and
$G$ is the gravitational constant.
Secondly, we have the equation for the binary mass function
\begin{equation}\label{mfn_eqn}
\frac{m_{1}q^3 \sin^3i}{(1+q)^2}=\frac{{K_1}^3 P}{2\pi G}~~,
\end{equation}
where $K_1$ is radial velocity semi-amplitude of the red giant and
$i$ is the angle of inclination of the orbit.
Next we use the relation between the 
light variation amplitude and the mass ratio for ellipsoidal variables given 
in \citet{2012MNRAS.423.2764N} 
\begin{equation}\label{iamp_eqn}
\frac{\Delta{I}}{0.87\sin^2i}=(0.221f^{4}+0.005)(1.44956q^{0.25}-0.44956)~~,
\end{equation}
where $\Delta{I}$ is the full light curve amplitude in the $I$ band and $f$ is 
the Roche lobe filling factor given by
\begin{equation}
	f=\frac{R_{1}}{R_{\rm{L}}}~~.
\end{equation}
Here, $R_{\rm{L}}$ is the effective Roche lobe radius \citep{1983ApJ...268..368E} 
given by 
\begin{equation}\label{egg_eqn}
	\frac{R_{\rm{L}}}{a}=\frac{0.49q^{-2/3}}{0.6q^{-2/3}+ln(1+q^{-1/3})}~,~0<q<\infty~~.
\end{equation}
The quantities $P$, $K_1$, $\Delta I$ and $R_1$ are 
known from observations of the light/velocity curves and from 2MASS
photometry, and we assume various values for the orbital inclination $i$
(see below).  In this situation, the five equations above 
can be solved iteratively for the five unknown
variables $m_1$, $q$, $a$, $f$ and $R_{\rm{L}}$ given that $i$ is prescribed.

As well as the above parameters, the WD code requires other input.
The systematic velocity $V_\gamma$ is determined by fitting the
velocity curve.  The zero-point of the orbital ephemeris HJD$_0$,
which corresponds to the light minimum where the radial velocity 
is becoming more negative (superior conjunction), is obtained by
fitting the $I$ band light curve to find the times of the two light minima
then selecting the one where the velocity curve is decreasing.  
The final object-related parameters that the WD code
requires as input are the effective temperatures $T_{\rm eff1}$ and
$T_{\rm eff2}$ of the red giant and companion, respectively, and the
dimensionless surface potentials $\Omega_1$ and $\Omega_2$ of these
stars (the WD code iterates to the best solution value of $\Omega_1$
but an input guess is required).  As noted above, $T_{\rm eff1}$ is
obtained from photometry while $\Omega_1$ is approximated by the
potential at outer end of star 1 along the line of centres where
\begin{equation}\label{om1_eqn}
   \Omega_1 = \frac{1}{\rho_1}+q(\frac{1}{1+\rho_1}+\rho_1)+\frac{1}{2}(1+q)\rho_1^2
\end{equation}
and $\rho_1 = R_1/a$.  The red giant radius $R_1$ is obtained from $T_{\rm eff1}$ 
and the observed luminosity $L_1$ using the equation
\begin{equation}\label{eq:bb}
   L_1 = 4\pi\sigma R_1^2 T_{\rm eff1}^4~~.
\end{equation}
To get the parameters for the secondary star, we assume that it lies 
in the middle of the main sequence. The evolutionary tracks of 
\citet{2008A&A...484..815B} for an LMC metallicity of $Z=0.008$ and 
masses between 0.8 and 8\msolar were approximated at this phase of 
evolution to yield the equations
\begin{equation}
   T_{\rm eff2}=6072 (\frac{m_2}{{\rm M}_{\odot}})^{0.599}
\end{equation}
and
\begin{equation}
   \frac{R_2}{{\rm R}_{\odot}}=1.16(\frac{m_2}{{\rm M}_{\odot}})^{0.756}
\end{equation}
where $R_2$ is the radius of the companion star.
The dimensionless surface potential $\Omega_2$ for the companion was then 
computed using the equation
\begin{equation}\label{eq:omega2}
   \Omega_2 = \frac{1}{\rho_2}+q_2(\frac{1}{1+\rho_2}+\rho_2)+\frac{1}{2}(1+q_2)\rho_2^2
\end{equation}
where $q_2=1/q$ and $\rho_2=R_2/a$.  Note that $L_2$ is not an input parameter
to the WD code but it is calculated using the analog of \autoref{eq:bb}.  
Provided the companion star is much less luminous than the red giant and of 
small size compared to the red giant Roche lobe radius, then the actual values of
$T_{\rm eff2}$ and $\Omega_2$ do not affect the orbital solution
significantly. 
In nearly all our objects, the companion
is indeed very faint compared to the red giant as it is not seen in either
the photometry or the spectra.  The companion is therefore most likely a
main sequence star or a compact star (white dwarf or neutron
star). However, we found no emission lines characteristic of an
accreting white dwarf or neutron star companion in any of our spectra
so main sequence star companions are most likely \citep{2014AJ....148..118N}.

\subsection{Creating an $i$-$q$ relation}

We are now in a position to use the WD code to create an $i$-$q$
relation for the system under study.
The WD code is composed of two main programs: DC and LC. The DC program
uses differential corrections to calculate the orbital parameters and
LC uses the converged solutions from DC to calculate stellar masses
and fit the light and velocity curves. The basic procedure is: for a
circular orbit binary system, we set HJD$_0$, $V_\gamma$, $P$, $T_{\rm
  eff1}$, $T_{\rm eff2}$, $\Omega_1$ and $\Omega_2$ with initial values derived as
described above, as well as setting $i = 90$\,degrees and $e = 0$.
The DC code is then set to iterate on HJD$_0$, $V_\gamma$, $a$,  $\Omega_1$ and $q$.
Then we step the input value of $i$ down from
90 degrees, in 5 degree increments to produce a one-dimensional family of
solutions (an $i$-$q$ sequence) for each system.  The solutions are found using mode 2 for
the ellipsoidal binaries, which are detached systems.  A simple
reflection treatment is used, with no spots, no proximity effects and
no third body in the system.  The stars are assumed to have their
rotation velocities synchronized at periastron.   
The gravity brightening exponent of the bolometric gravity
brightening law for the red giant, $g_1$, is set to the value given by
the non-grey version of Equation~2 in \citet{1997A&A...326..257A} who evaluated $g_1$
from convective stellar atmosphere models.  For our typical red giants
with $T_{\rm eff1} \sim 4000$\,K, $g_1\approx 0.3$.  For the main
sequence companion, we set $g_2$=1.0 as appropriate for a radiative
envelope, although in practice the value of $g_2$ is unimportant due
to the relatively low luminosity of the companion to the red giant.
Bolometric albedos for reflection heating and
re-radiation are set to A$_1$=0.5 for the red giant and A$_2$=1.0 for
the companion, again as expected for convective and radiative
envelopes.  Limb darkening is treated using the bolometric square root
law, with the coefficients determined locally in the DC code.

In general, DC solutions converged (corrections $\ll$ errors) in six
to eight iterations when $i$ was set.  Stepping of $i$ was 
stopped at 30 degrees, or when the solution was deemed 
to be getting too poor. Solutions
with either companion exceeding its critical Roche lobe radius were
rejected.  For eccentric systems, a similar approach was applied
but DC was allowed to adjust two extra parameters, the eccentricity
$e$ and the argument of periastron for the red giant, $\omega$. 

\subsection{Finding the best solution from the $i$-$q$ relation}\label{best_sol}

\citet{1985ApJ...295..143M} showed that the time variation of
the light curves of ellipsoidal variables could be approximated by the
sum of terms proportional to $\cos$($\phi$), $\cos$(2$\phi$) and
$\cos$(3$\phi$) where $\phi$ is the orbital phase.  The coefficients
of these terms depend on the orbital parameters $q$, $\sin$($i$) and
$R_1/a$ as well as on gravity brightening and limb darkening.  The
dominant $\cos$(2$\phi$) term gives rise to the characteristic light
curve shape with two cycles of the light curve per orbital period and
this term essentially determines the total amplitude of the
ellipsoidal light variation.  Adjusting the orbital parameters to fit
the $\cos$(2$\phi$) term to the light curve is equivalent to defining
\autoref{iamp_eqn}.  As seen above, such fitting of the total
amplitude of light variation is sufficient to give an initial guess for 
the orbital parameters of ellipsoidal variables when $i$ is pre-defined.  
However, further information is required to estimate $i$ as well as the
other orbital parameters.  This information comes largely from the
$\cos$(3$\phi$) term.

The $\cos$(3$\phi$) term gives rise to alternating depths for the two
minima in the light curve per orbital period.  Such alternating depths
are a characteristic of the light curves of ellipsoidal variables.  By
fitting both the total amplitude of variation and the differing depths
of alternate minima in the light curve, two independent constraints
are placed on the orbital parameters.  In combination with Equations
\ref{kep_eqn} and \ref{mfn_eqn}, there are four constraints which
allow a complete determination of the orbital parameters $a$,
$\Omega_1$, $q$ and $i$.  In practice, when using the WD code, $K_1$
in \autoref{mfn_eqn} comes from fitting the velocity curve, $R_1$ is
obtained from \autoref{eq:bb} since $T_{\rm eff1}$ is known,
$\Omega_1$ comes from an accurate version of \autoref{om1_eqn} and the
other solution parameters $V_\gamma$ and HJD$_0$ come from fits to the
velocity and light curves of the red giant.

Our procedure for obtaining a complete orbital solution, including
the determination of $i$ and $q$, was as follows.  Firstly, we
obtained the $i$-$q$ sequence for each object as described in the last
sub-section.  To obtain the final solution, which involves iteration for $i$ and $q$, 
a reasonably reliable set of starting parameters was
required and these were obtained from that solution belonging to the
$i$-$q$ sequence that had the smallest $\Sigma r^2$, 
where $\Sigma r^2$ is the sum of squares of
residuals from the light curve, as given in the DC output.
Sometimes the plot of $\Sigma r^2$ against $i$ showed two minima, one
at or near $i = 90$ degrees and one at a smaller value of $i$.  The starting
parameters of the deepest minimum were adopted unless $m_2 > m_1$
for this minimum in which case the solution is unlikely since
the lack of evidence for white dwarf or neutron star companions
\citep{2014AJ....148..118N} suggests that the red giant is the primary star in all cases.
When $m_2 > m_1$ for the deeper minimum, the starting parameters of the
second minimum were adopted (this is always the minimum at or near $i = 90$ degrees).  
Given the starting parameters, we then iterated on the full
set of orbital parameters HJD$_0$, $V_\gamma$, $i$, $a$,  $\Omega_1$ and 
$q$ until a fully converged model was obtained.
The converged results from DC were put into LC to calculate
the light and velocity curves, as well as $m_1$, $m_2$, $R_1$, $R_2$,
$L_1$ and $L_2$. As part of its solution process, DC calculates
the standard errors for the parameters HJD$_0$, $V_\gamma$, $i$, $a$,  $\Omega_1$
and $q$.  These standard errors can then be used to compute the
standard errors for $m_1$ and $m_2$ via \autoref{kep_eqn} and the definition of $q$.

In some cases, the $i$-$q$ relation indicated that the best
fit model should have $i = 90$ degrees or very close to it.  The WD code
can not iterate to the best $i$ value when it is 90 degrees since the 
solution is independent of $i$ in this situation.  When the WD code failed
to converge for these models, we adopted $i = 90$ degrees and iterated 
the remaining parameters to the best solution.  Although the WD code can
not estimate an error for $i$ in these cases, examination of 
$i$-$q$ relation for these systems suggests that the error should be less than
10 degrees.

We found two objects that had $q \ga 1$ but with the $q$ value
less than 1-sigma above unity.  OGLE 053438.78 -695634.1 has 
$q = 1.047\pm0.218$ and $R_1/R_{L,1} = 0.96$ so this might be a case where
stable mass transfer is occurring onto a main sequence star without
prominent emission lines in the spectrum of the object.  
The other object, OGLE 052853.08 -695557.9, has $q = 1.058\pm0.130$ but
$R_1/R_{L,1} = 0.83$ so significant mass transfer is unlikely.
A problem with the solutions for these objects is that it is not possible
to know the contribution of the secondary star to the light curve.
We have assumed that this contribution is negligible based on the fact that
we do not see multiple lines with different velocities in the spectra.
In the case of OGLE 052853.08 -695557.9, the lack of a bright secondary 
implies that $q$ is actually slightly less than unity.

\capstartfalse
\tabletypesize{\tiny}
\begin{center}
\begin{longtable*}{cccccccccccccccccc}
\caption{Orbital solution for each ellipsoidal variable} \label{tb:best}\\
\hline
\multicolumn{1}{c}{$Object$}&
\multicolumn{1}{c}{$P$}&
\multicolumn{1}{c}{$T_{\rm eff1}$}&
\multicolumn{1}{c}{$HJD_0$}&
\multicolumn{1}{c}{$V_\gamma$}&
\multicolumn{1}{c}{$i$}&
\multicolumn{1}{c}{$a$}&    \multicolumn{1}{c}{$e$}&    \multicolumn{1}{c}{$\omega$}&
\multicolumn{1}{c}{$q$}&    \multicolumn{1}{c}{$\Omega_1$}&
\multicolumn{1}{c}{$m_1$}&  \multicolumn{1}{c}{$m_2$}&
\multicolumn{1}{c}{$R_1$}&  \multicolumn{1}{c}{$R_{L,1}$}&  \multicolumn{1}{c}{$R_2$}& 
\multicolumn{1}{c}{$L_1$}& \multicolumn{1}{c}{$L_2$}
\\
\multicolumn{1}{c}{OGLE II}&
\multicolumn{1}{c}{day}&
\multicolumn{1}{c}{(K)}&
\multicolumn{1}{c}{(2450000.0+)}&
\multicolumn{1}{c}{\kms}&
\multicolumn{1}{c}{(deg)}&  
\multicolumn{1}{c}{($\rm{R_{\odot}}$)}&    \multicolumn{1}{c}{}&    \multicolumn{1}{c}{(rad)}&
\multicolumn{1}{c}{}&   \multicolumn{1}{c}{}&
\multicolumn{1}{c}{$(\rm{M_{\odot}})$}&  \multicolumn{1}{c}{$(\rm{M_{\odot}})$}&
\multicolumn{1}{c}{($\rm{R_{\odot}}$)}& \multicolumn{1}{c}{($\rm{R_{\odot}}$)}&  \multicolumn{1}{c}{($\rm{R_{\odot}}$)}&
\multicolumn{1}{c}{($\rm{L_{\odot}}$)}& \multicolumn{1}{c}{($\rm{L_{\odot}}$)}   
\\
\hline
\endfirsthead

\multicolumn{18}{c}
{{ \tablename\ \thetable{} (continued) }} \\
\hline
\multicolumn{1}{c}{$Object$}&
\multicolumn{1}{c}{$P$}& 
\multicolumn{1}{c}{$T_{\rm eff1}$}&
\multicolumn{1}{c}{$HJD_0$}&
\multicolumn{1}{c}{$V_\gamma$}&
\multicolumn{1}{c}{$i$}&
\multicolumn{1}{c}{$a$}&    \multicolumn{1}{c}{$e$}&    \multicolumn{1}{c}{$\omega$}&
\multicolumn{1}{c}{$q$}&    \multicolumn{1}{c}{$\Omega_1$}&
\multicolumn{1}{c}{$m_1$}&  \multicolumn{1}{c}{$m_2$}&
\multicolumn{1}{c}{$R_1$}&  \multicolumn{1}{c}{$R_{L,1}$}& \multicolumn{1}{c}{$R_2$}&
\multicolumn{1}{c}{$L_1$}&  \multicolumn{1}{c}{$L_2$}
\\
\multicolumn{1}{c}{OGLE II}& 
\multicolumn{1}{c}{(day)}&
\multicolumn{1}{c}{(K)}&
\multicolumn{1}{c}{(2450000.0+)}&
\multicolumn{1}{c}{(\kms)}&
\multicolumn{1}{c}{(deg)}& 
\multicolumn{1}{c}{($\rm{R_{\odot}}$)}&    \multicolumn{1}{c}{}&    \multicolumn{1}{c}{(rad)}&
\multicolumn{1}{c}{}&   \multicolumn{1}{c}{}&
\multicolumn{1}{c}{$(\rm{M_{\odot}})$}&  \multicolumn{1}{c}{$(\rm{M_{\odot}})$}&
\multicolumn{1}{c}{($\rm{R_{\odot}}$)}&  \multicolumn{1}{c}{($\rm{R_{\odot}}$)}& \multicolumn{1}{c}{($\rm{R_{\odot}}$)}&  
\multicolumn{1}{c}{($\rm{L_{\odot}}$)}&  \multicolumn{1}{c}{($\rm{L_{\odot}}$)}
\\
\hline
\endhead

\hline
\endfoot
\hline
\endlastfoot
\multicolumn{18}{c}{\textbf{Circular orbits~~(\citet{2014AJ....148..118N})} } \\
\hline
050222.40-691733.6&  107.8& 3990& 5628.0$\pm$0.4& 256.6$\pm$0.4& 69.8$\pm$19.5& 136.6$\pm$4.1 &   0.000& -& 0.377$\pm$0.078& 3.093$\pm$0.090& 2.14$\pm$0.23& 0.81$\pm$0.19&  51.6& 63.4& 0.7&  560.7&  0.4 \\
050254.15-692013.8&  171.4& 4234& 5561.0$\pm$0.6& 262.1$\pm$0.3& 52.9$\pm$29.2& 198.3$\pm$11.1&   0.000& -& 0.553$\pm$0.367& 4.493$\pm$0.286& 2.30$\pm$0.66& 1.27$\pm$0.58&  50.8& 85.5& 1.3&  747.1&  3.7 \\
050258.71-684406.6&  433.8& 4020& 5917.5$\pm$0.9& 255.6$\pm$0.2& 90.0~~~~~~~~~& 354.8$\pm$5.6 &   0.000& -& 0.905$\pm$0.030& 5.938$\pm$0.068& 1.67$\pm$0.08& 1.52$\pm$0.08&  70.9&137.5& 2.0& 1103.4& 10.9 \\ 
050350.55-691430.2&  224.7& 4030& 5553.5$\pm$0.6& 221.7$\pm$0.2& 54.7$\pm$7.6 & 192.8$\pm$3.3 &   0.000& -& 0.476$\pm$0.075& 3.635$\pm$0.069& 1.29$\pm$0.09& 0.61$\pm$0.07&  62.1& 85.7& 0.6&  857.8&  0.1 \\
050353.41-690230.8&  323.6& 4049& 5829.3$\pm$1.1& 243.4$\pm$0.5& 43.0$\pm$3.7 & 172.7$\pm$11.4&   0.000& -& 0.150$\pm$0.036& 2.209$\pm$0.140& 0.58$\pm$0.11& 0.09$\pm$0.02&  88.6& 94.2& 0.0& 1799.4&  0.0 \\
050438.97-693115.3&  383.2& 4312& 5721.1$\pm$1.5& 239.8$\pm$0.4& 50.7$\pm$9.8 & 212.8$\pm$7.5 &   0.000& -& 0.464$\pm$0.115& 3.425$\pm$0.131& 0.60$\pm$0.08& 0.28$\pm$0.06&  73.4& 95.1& 0.0& 1740.1&  0.0 \\
050504.70-683340.3&  238.9& 4327& 5914.0$\pm$0.7& 247.5$\pm$0.7& 86.2$\pm$9.7 & 220.7$\pm$5.0 &   0.000& -& 0.917$\pm$0.059& 5.866$\pm$0.158& 1.32$\pm$0.10& 1.21$\pm$0.09&  44.8& 85.3& 1.5&  651.8&  3.9 \\ 
050512.19-693543.5& 204.3&  3988& 5822.2$\pm$1.1& 313.4$\pm$0.4& 75.7$\pm$11.9& 203.6$\pm$8.6 &   0.000& -& 0.681$\pm$0.094& 4.732$\pm$0.118& 1.62$\pm$0.23& 1.10$\pm$0.17& 50.7 & 84.0& 1.1&  540.3&  2.1 \\
050554.57-683428.5& 270.1&  4061& 5896.9$\pm$0.6& 238.8$\pm$0.4& 37.8$\pm$2.3 & 363.4$\pm$5.1 &   0.000& -& 0.445$\pm$0.042& 3.322$\pm$0.059& 6.11$\pm$0.31& 2.72$\pm$0.21& 129.2&163.6& 1.7& 3874.9& 51.3 \\
050558.70-682208.9& 264.0&  4128& 5771.1$\pm$1.1& 260.7$\pm$0.2& 87.0$\pm$10.0& 260.7$\pm$7.8 &   0.000& -& 0.295$\pm$0.013& 4.607$\pm$0.119& 2.64$\pm$0.24& 0.78$\pm$0.07& 60.8 &126.9& 0.7&  933.2&  0.4 \\  
050604.99-681654.9& 201.7&  4174& 5579.1$\pm$0.7& 286.6$\pm$0.3& 54.6$\pm$30.1& 215.4$\pm$11.2&   0.000& -& 0.516$\pm$0.324& 4.487$\pm$0.249& 2.18$\pm$0.58& 1.12$\pm$0.50& 54.7 & 94.2& 1.1&  802.1&  2.0 \\
050651.36-695245.4& 220.2&  4192& 5832.9$\pm$0.7& 229.1$\pm$0.2& 50.4$\pm$16.5& 187.3$\pm$4.2 &   0.000& -& 0.860$\pm$0.399& 4.808$\pm$0.403& 0.98$\pm$0.22& 0.84$\pm$0.22& 48.0 & 73.4& 0.9&  630.7&  0.8 \\
050659.79-692540.4& 156.4&  4254& 5758.8$\pm$0.2& 259.6$\pm$0.2& 86.4$\pm$ 5.7& 194.7$\pm$1.4 &   0.000& -& 0.675$\pm$0.013& 4.083$\pm$0.025& 2.42$\pm$0.05& 1.63$\pm$0.04& 58.0 & 80.5& 0.9& 1000.7&  0.8 \\
050709.66-683824.8& 206.1&  4121& 5710.2$\pm$0.6& 235.7$\pm$0.3& 57.0$\pm$11.8& 226.6$\pm$4.3 &   0.000& -& 0.651$\pm$0.157& 3.959$\pm$0.153& 2.23$\pm$0.25& 1.45$\pm$0.23& 69.7 & 94.4& 1.5& 1215.9&  9.6 \\
050720.82-683355.2& 521.4&  4072& 5791.3$\pm$2.0& 247.5$\pm$0.3& 88.0$\pm$10.2& 600.2$\pm$9.7 &   0.000& -& 0.346$\pm$0.011& 5.169$\pm$0.075& 7.94$\pm$0.39& 2.75$\pm$0.15& 125.0&283.8& 1.9& 3670.5& 99.6 \\
050843.38-692815.1& 121.8&  4098& 5643.0$\pm$0.5& 249.9$\pm$0.2& 46.1$\pm$13.9& 123.4$\pm$4.9 &   0.000& -& 0.461$\pm$0.178& 3.402$\pm$0.160& 1.16$\pm$0.20& 0.54$\pm$0.16& 42.8 & 55.2& 0.5&  446.3&  0.1 \\ 
050900.02-690427.2& 156.5&  4273& 5628.6$\pm$1.0& 266.1$\pm$0.2& 48.3$\pm$29.2& 140.6$\pm$12.1&   0.000& -& 0.324$\pm$0.226& 3.621$\pm$0.147& 1.15$\pm$0.36& 0.37$\pm$0.22& 43.2 & 67.3& 0.3&  567.5&  0.0 \\            
050948.63-690157.4&  75.6&  4529& 5598.6$\pm$0.1& 253.6$\pm$0.4& 90.0~~~~~~~~~& 139.9$\pm$1.0 &   0.000& -& 0.204$\pm$0.007& 2.277$\pm$0.021& 5.35$\pm$0.12& 1.09$\pm$0.04& 71.5 & 72.6& 1.0& 2080.1&  1.6 \\  
051050.92-692228.0& 326.2&  4284& 5789.3$\pm$1.7& 263.2$\pm$0.3& 38.9$\pm$ 8.9& 193.7$\pm$6.5 &   0.000& -& 0.585$\pm$0.204& 3.831$\pm$0.197& 0.58$\pm$0.09& 0.34$\pm$0.08& 60.7 & 82.5& 0.3& 1138.3&  0.0 \\
051205.43-684559.9& 390.5&  3912& 5884.7$\pm$1.6& 280.3$\pm$0.4& 68.9$\pm$29.8& 296.4$\pm$11.0&   0.000& -& 0.557$\pm$0.203& 4.169$\pm$0.166& 1.47$\pm$0.25& 0.82$\pm$0.21& 83.0 &127.6& 0.8& 1300.4&  0.5 \\
051256.36-684937.5& 341.9&  3784& 6009.8$\pm$0.5& 241.6$\pm$0.5& 64.9$\pm$ 6.7& 291.0$\pm$5.3 &   0.000& -& 0.611$\pm$0.068& 3.483$\pm$0.092& 1.76$\pm$0.12& 1.08$\pm$0.09& 104.1&122.8& 1.1& 1659.9&  1.8 \\
051345.17-692212.1& 197.9&  4038& 5694.2$\pm$0.8& 255.7$\pm$0.3& 76.7$\pm$12.1& 202.8$\pm$11.9&   0.000& -& 0.741$\pm$0.112& 5.344$\pm$0.196& 1.64$\pm$0.31& 1.22$\pm$0.24&  44.3& 82.1& 1.4&  441.5&  3.6 \\
051620.47-690755.4& 240.7&  3963& 5758.1$\pm$0.3& 229.3$\pm$0.3& 63.9$\pm$ 4.2& 214.1$\pm$1.9 &   0.000& -& 0.901$\pm$0.067& 3.770$\pm$0.085& 1.20$\pm$0.05& 1.08$\pm$0.05& 77.2 & 83.1& 1.4& 1212.3&  2.7 \\
051621.06-692929.6& 179.3&  4294& 5540.3$\pm$0.7& 264.9$\pm$0.4& 39.9$\pm$ 4.3& 252.4$\pm$6.8 &   0.000& -& 0.221$\pm$0.032& 2.943$\pm$0.066& 5.51$\pm$0.47& 1.21$\pm$0.17& 94.8 &129.3& 1.2& 2803.2&  2.5 \\
051653.08-690651.2& 269.2&  3792& 5801.8$\pm$0.4& 264.1$\pm$0.4& 62.0$\pm$ 3.5& 285.5$\pm$3.6 &   0.000& -& 0.415$\pm$0.026& 3.027$\pm$0.045& 3.05$\pm$0.13& 1.26$\pm$0.07& 112.7&130.4& 1.3& 1970.1&  3.8 \\
051746.55-691750.2& 225.2&  3985& 5881.6$\pm$0.4& 270.7$\pm$0.3& 43.5$\pm$ 2.2& 243.7$\pm$4.1 &   0.000& -& 0.255$\pm$0.020& 2.659$\pm$0.044& 3.05$\pm$0.16& 0.78$\pm$0.06& 105.0&121.8& 0.7& 2310.1&  0.4 \\
052012.26-694417.5& 177.4&  3983& 5649.1$\pm$0.4& 254.9$\pm$0.2& 42.0$\pm$ 2.5& 215.7$\pm$2.9 &   0.000& -& 0.260$\pm$0.020& 2.827$\pm$0.029& 3.40$\pm$0.15& 0.88$\pm$0.07& 86.4 &107.4& 0.8& 1559.6&  0.6 \\
052029.89-695934.1& 171.0&  4492& 5640.8$\pm$0.5& 240.3$\pm$0.3& 82.6$\pm$13.3& 254.3$\pm$ 3.6&   0.000& -& 0.484$\pm$0.028& 4.479$\pm$0.048& 5.09$\pm$0.23& 2.46$\pm$0.14& 64.2 &112.6& 1.8& 1603.2& 54.6 \\
052115.05-693155.2& 195.3&  4525& 5661.5$\pm$0.4& 267.0$\pm$0.3& 78.2$\pm$ 0.8& 290.9$\pm$4.2 &   0.000& -& 0.506$\pm$0.014& 5.546$\pm$0.066& 5.76$\pm$0.26& 2.91$\pm$0.14& 57.9 &127.7& 2.0& 1352.0&105.9 \\
052117.49-693124.8& 243.3&  3908& 5605.6$\pm$0.4& 234.0$\pm$0.2& 63.9$\pm$ 8.0& 242.7$\pm$2.6 &   0.000& -& 0.706$\pm$0.093& 4.128$\pm$0.088& 1.90$\pm$0.12& 1.34$\pm$0.11& 72.0 & 99.3& 1.4&  971.8&  5.4 \\
052119.55-710022.1& 300.9&  3907& 5983.3$\pm$0.5& 265.4$\pm$0.3& 47.3$\pm$ 2.9& 233.2$\pm$3.3 &   0.000& -& 0.701$\pm$0.068& 3.490$\pm$0.091& 1.11$\pm$0.06& 0.77$\pm$0.06& 86.2 & 95.7& 0.7& 1394.4&  0.4 \\
052238.44-691715.1&  77.5&  5046& 5596.2$\pm$0.1& 259.4$\pm$0.3& 54.9$\pm$ 3.5& 144.9$\pm$1.3 &   0.000& -& 0.537$\pm$0.041& 3.387$\pm$0.048& 4.42$\pm$0.16& 2.37$\pm$0.13& 52.1 & 62.8& 1.6& 1782.0& 33.1 \\
052324.57-692924.2& 312.4&  4388& 5900.6$\pm$0.9& 290.5$\pm$0.4& 85.2$\pm$ 4.4& 454.2$\pm$8.0 &   0.000& -& 0.383$\pm$0.014& 5.149$\pm$0.080& 9.32$\pm$0.50& 3.57$\pm$0.21& 95.7 &210.6& 2.2& 3184.4&186.1 \\
052422.28-692456.2& 485.4&  3948& 5996.3$\pm$2.2& 284.8$\pm$0.5& 90.0~~~~~~~~~& 291.2$\pm$7.7 &   0.000& -& 0.442$\pm$0.033& 3.880$\pm$0.115& 0.98$\pm$0.08& 0.43$\pm$0.04& 85.8 &131.3& 0.4& 1464.7&  0.0 \\ 
052425.52-695135.2& 206.7&  4345& 5591.3$\pm$0.5& 256.8$\pm$0.3& 48.0$\pm$ 6.9& 267.0$\pm$3.4 &   0.000& -& 0.721$\pm$0.142& 4.346$\pm$0.152& 3.48$\pm$0.32& 2.51$\pm$0.30& 74.7 &108.9& 2.2& 1849.5& 66.8 \\
052438.18-700435.9& 192.8&  4314& 5753.5$\pm$0.5& 231.3$\pm$0.2& 64.7$\pm$19.5& 188.6$\pm$5.1 &   0.000& -& 0.409$\pm$0.105& 4.164$\pm$0.071& 1.72$\pm$0.19& 0.70$\pm$0.14& 50.7 & 84.6& 0.6&  820.9&  0.2 \\
052510.82-700124.0& 189.7&  4328& 5761.1$\pm$0.4& 241.5$\pm$0.2& 59.0$\pm$13.4& 276.5$\pm$6.7 &   0.000& -& 0.314$\pm$0.067& 4.101$\pm$0.049& 6.01$\pm$0.53& 1.89$\pm$0.33& 73.6 &133.1& 1.5& 1760.2& 18.0 \\
052554.91-694137.4& 182.8&  4008& 5760.0$\pm$0.1& 244.7$\pm$0.2& 74.5$\pm$ 4.1& 201.4$\pm$1.4 &   0.000& -& 0.628$\pm$0.025& 3.541$\pm$0.036& 2.02$\pm$0.05& 1.27$\pm$0.04& 70.9 & 84.6& 1.3& 1087.5&  3.9 \\
052703.64-694837.8& 350.4&  4131& 5629.2$\pm$0.3& 250.5$\pm$0.7& 76.1$\pm$ 0.9& 483.2$\pm$4.7 &   0.000& -& 0.665$\pm$0.020& 3.765$\pm$0.049& 7.41$\pm$0.23& 4.93$\pm$0.17& 159.1&200.4& 3.1& 6433.0& 807.0\\
052833.50-695834.6& 183.2&  4186& 5656.2$\pm$0.2& 249.4$\pm$0.4& 55.9$\pm$ 2.6& 237.9$\pm$5.0 &   0.000& -& 0.236$\pm$0.019& 2.618$\pm$0.058& 4.36$\pm$0.28& 1.03$\pm$0.09& 103.5&120.5& 0.9& 2918.4&  1.0 \\
052954.82-700622.5& 154.7&  4149& 5642.0$\pm$0.6& 274.4$\pm$0.3& 90.0~~~~~~~~~& 171.7$\pm$10.1&   0.000& -& 0.584$\pm$0.056& 4.815$\pm$0.197& 1.79$\pm$0.32& 1.05$\pm$0.20& 40.9 & 73.2& 1.0&  433.1&  1.5 \\
053337.07-703111.7& 315.5&  3799& 5726.2$\pm$0.5& 243.9$\pm$0.6& 56.4$\pm$ 3.4& 229.8$\pm$7.2 &   0.000& -& 0.298$\pm$0.029& 2.513$\pm$0.081& 1.26$\pm$0.12& 0.38$\pm$0.05& 109.1&111.7& 0.3& 1883.2&  0.0 \\
053338.94-694455.2& 126.9&  4128& 5621.2$\pm$0.2& 239.0$\pm$1.0& 90.0~~~~~~~~~& 122.3$\pm$3.1 &   0.000& -& 0.461$\pm$0.040& 3.283$\pm$0.110& 1.04$\pm$0.09& 0.48$\pm$0.05& 44.4 & 54.7& 0.4&  498.6&  0.1 \\ 
053356.79-701919.6& 205.3&  4169& 5639.7$\pm$0.6& 268.4$\pm$0.3& 80.7$\pm$22.1& 192.3$\pm$8.5 &   0.000& -& 0.777$\pm$0.120& 5.578$\pm$0.163& 1.28$\pm$0.19& 0.99$\pm$0.16& 40.3 & 77.1& 1.1&  431.7&  1.5 \\
053438.78-695634.1& 384.7&  3751& 5878.2$\pm$0.9& 213.4$\pm$0.5& 27.9$\pm$ 2.1& 308.2$\pm$10.8&   0.000& -& 1.047$\pm$0.218& 3.931$\pm$0.320& 1.30$\pm$0.19& 1.36$\pm$0.20& 110.8&115.6& 0.0& 1784.3&  0.0 \\ 
053733.22-695026.8& 263.0&  3970& 5786.8$\pm$0.7& 316.9$\pm$0.2& 49.1$\pm$ 4.7& 216.9$\pm$3.2 &   0.000& -& 0.456$\pm$0.055& 3.397$\pm$0.056& 1.36$\pm$0.08& 0.62$\pm$0.06& 75.3 & 97.2& 0.6& 1163.6&  0.1 \\
053946.88-704257.8& 265.0&  3929& 5987.2$\pm$0.7& 227.6$\pm$0.3& 67.0$\pm$43.8& 234.3$\pm$5.9 &   0.000& -& 0.979$\pm$0.638& 5.613$\pm$0.691& 1.24$\pm$0.41& 1.22$\pm$0.41& 50.9 & 89.2& 2.9&  500.7&  5.3 \\
054258.34-701609.2& 164.0&  4114& 5578.8$\pm$0.9& 298.5$\pm$0.2& 90.0~~~~~~~~~& 177.1$\pm$6.5 &   0.000& -& 0.223$\pm$0.011& 3.601$\pm$0.114& 2.27$\pm$0.25& 0.51$\pm$0.06& 53.0 & 90.6& 0.4&  696.9&  0.1 \\
054736.16-705627.2& 135.9&  4244& 5624.8$\pm$0.3& 277.9$\pm$0.2& 84.2$\pm$ 9.0& 188.0$\pm$5.0 &   0.000& -& 0.334$\pm$0.016& 4.297$\pm$0.093& 3.62$\pm$0.29& 1.21$\pm$0.11& 47.8 & 89.5& 1.2&  671.3&  2.5 \\
\hline                                                                                                                                                                                   
\multicolumn{18}{c}{\textbf{Circular orbits~~(\citet{2010MNRAS.405.1770N})} } \\                                                                                                         
\hline
052613.43-694448.2& 110.5&  4027& 5595.1$\pm$0.1& 281.6$\pm$0.2& 90.0~~~~~~~~~& 143.7$\pm$0.7 &   0.000& -& 0.687$\pm$0.009& 3.602$\pm$0.020& 1.94$\pm$0.03& 1.33$\pm$0.02& 50.6 & 59.2& 1.4&  567.4& 4.9  \\
052656.70-695204.0& 442.7&  4046& 5609.3$\pm$1.7& 247.5$\pm$0.5& 90.0~~~~~~~~~& 293.3$\pm$7.9 &   0.000& -& 0.301$\pm$0.024& 3.503$\pm$0.105& 1.33$\pm$0.11& 0.40$\pm$0.04& 92.9 &142.3& 0.4& 1963.9& 0.0  \\
052740.75-693539.5& 131.4&  4407& 5639.1$\pm$0.1& 252.1$\pm$0.1& 80.5$\pm$ 5.9& 139.2$\pm$0.5 &   0.000& -& 0.832$\pm$0.029& 4.117$\pm$0.033& 1.15$\pm$0.02& 0.95$\pm$0.02& 43.2 & 55.0& 1.0&  663.5& 1.3  \\
052805.59-694732.8& 139.2&  4194& 5603.6$\pm$0.1& 350.8$\pm$0.1& 71.7$\pm$ 4.6& 153.3$\pm$0.7 &   0.000& -& 0.680$\pm$0.034& 3.644$\pm$0.036& 1.49$\pm$0.04& 1.01$\pm$0.03& 53.0 & 63.2& 1.0&  773.7& 1.4  \\
052808.39-695504.9& 157.7&  3973& 5659.0$\pm$0.2& 255.2$\pm$0.2& 90.0~~~~~~~~~& 141.3$\pm$0.7 &   0.000& -& 0.418$\pm$0.007& 2.972$\pm$0.018& 1.08$\pm$0.02& 0.45$\pm$0.01& 57.2 & 64.4& 0.4&  675.0& 0.0  \\
052837.68-695250.9& 327.4&  3806& 5793.3$\pm$0.6& 275.0$\pm$0.4& 74.0$\pm$14.5& 291.2$\pm$5.2 &   0.000& -& 0.539$\pm$0.074& 3.534$\pm$0.081& 2.01$\pm$0.15& 1.08$\pm$0.11& 99.3 &126.2& 1.1& 1569.8& 1.7  \\
052852.69-694933.6& 128.8&  4304& 5603.5$\pm$0.2& 231.1$\pm$0.1& 71.0$\pm$ 9.2& 139.6$\pm$1.4 &   0.000& -& 0.396$\pm$0.035& 3.436$\pm$0.028& 1.58$\pm$0.06& 0.62$\pm$0.04& 46.7 & 64.3& 0.6&  689.5& 0.1  \\
052853.08-695557.9& 191.5&  4332& 5609.1$\pm$0.3& 238.2$\pm$0.2& 40.9$\pm$ 2.9& 231.0$\pm$1.5 &   0.000& -& 1.058$\pm$0.130& 4.339$\pm$0.154& 2.19$\pm$0.15& 2.32$\pm$0.15& 72.0 & 86.4& 0.0& 1694.2& 0.0  \\
052913.51-695632.3&  65.7&  4804& 5580.0$\pm$0.1& 255.0$\pm$0.3& 69.4$\pm$ 0.5& 134.9$\pm$0.7 &   0.000& -& 0.778$\pm$0.014& 3.730$\pm$0.023& 4.30$\pm$0.08& 3.35$\pm$0.06& 47.0 & 54.1& 2.6& 1169.6&230.2 \\
052917.67-694137.9& 282.5&  3918& 5673.5$\pm$0.4& 305.1$\pm$0.2& 75.5$\pm$ 9.9& 228.2$\pm$1.6 &   0.000& -& 0.681$\pm$0.056& 3.874$\pm$0.064& 1.19$\pm$0.05& 0.81$\pm$0.04& 72.9 & 94.1& 0.8& 1010.6& 0.5  \\
\hline
\multicolumn{18}{c}{\textbf{Eccentric orbits~~(\citet{2014AJ....148..118N})} } \\
\hline
050610.03-683153.0& 318.5&  4197& 5766.7$\pm$1.3& 279.3$\pm$0.3& 61.8$\pm$16.2& 278.0$\pm$ 9.0& 0.202$\pm$0.018& 0.327$\pm$0.124& 0.830$\pm$0.229&  6.242$\pm$0.350& 1.56$\pm$0.25& 1.29$\pm$0.23&  53.7& 109.9& 1.6&  795.3&  5.1 \\ 
051009.20-690020.0& 321.3&  4023& 5997.8$\pm$0.0& 292.1$\pm$0.4& 74.3$\pm$22.6& 264.0$\pm$10.0& 0.090$\pm$0.021& 2.191$\pm$0.186& 0.299$\pm$0.055&  4.225$\pm$0.113& 1.84$\pm$0.22& 0.55$\pm$0.10&  68.5& 128.2& 0.5& 1035.0&  0.1 \\
051101.04-691425.1& 282.5&  4173& 5809.2$\pm$2.0& 212.7$\pm$0.6& 55.9$\pm$ 4.7& 217.9$\pm$ 7.5& 0.254$\pm$0.022& 6.432$\pm$0.108& 0.416$\pm$0.053&  4.814$\pm$0.179& 1.23$\pm$0.13& 0.51$\pm$0.07&  52.3&  99.4& 0.5&  730.9&  0.1 \\
051200.23-690838.4& 641.3&  3737& 6504.3$\pm$2.8& 245.6$\pm$0.4& 90.0~~~~~~~~~& 455.3$\pm$12.5& 0.278$\pm$0.013& 3.824$\pm$0.068& 0.237$\pm$0.017&  3.637$\pm$0.109& 2.49$\pm$0.21& 0.59$\pm$0.06& 139.3& 230.4& 0.5& 2730.9&  0.1 \\
051220.63-684957.8& 486.5&  3983& 5944.3$\pm$1.9& 308.7$\pm$0.4& 90.0~~~~~~~~~& 338.1$\pm$ 7.0& 0.061$\pm$0.014& 6.267$\pm$0.218& 0.379$\pm$0.023&  3.798$\pm$0.089& 1.59$\pm$0.10& 0.60$\pm$0.05& 100.9& 157.1& 0.5& 2127.2&  0.1 \\
051347.73-693049.7& 306.6&  4114& 5773.3$\pm$1.2& 252.5$\pm$0.3& 36.1$\pm$ 3.4& 214.8$\pm$ 4.4& 0.138$\pm$0.020& 4.578$\pm$0.113& 0.837$\pm$0.143&  4.721$\pm$0.190& 0.77$\pm$0.08& 0.65$\pm$0.07&  58.7&  84.7& 0.6&  856.2&  0.2 \\
051738.19-694848.4& 297.7&  3943& 5882.7$\pm$1.3& 238.5$\pm$0.3& 74.5$\pm$15.3& 275.5$\pm$ 6.3& 0.353$\pm$0.017& 5.354$\pm$0.060& 0.443$\pm$0.057&  4.744$\pm$0.126& 2.20$\pm$0.17& 0.97$\pm$0.11&  68.4& 124.2& 0.9&  923.3&  1.0 \\ 
051818.84-690751.3& 463.9&  3710& 6119.3$\pm$1.3& 264.0$\pm$0.2& 90.0~~~~~~~~~& 336.5$\pm$ 5.8& 0.195$\pm$0.014& 4.655$\pm$0.075& 0.528$\pm$0.021&  4.753$\pm$0.073& 1.56$\pm$0.08& 0.82$\pm$0.05&  82.8& 146.4& 0.8&  909.1&  0.4 \\ 
052542.11-694847.4& 514.6&  4004& 5703.5$\pm$1.9& 263.8$\pm$0.7& 54.5$\pm$ 1.5& 473.9$\pm$10.2& 0.267$\pm$0.008& 1.006$\pm$0.049& 0.696$\pm$0.053&  5.553$\pm$0.156& 3.18$\pm$0.23& 2.22$\pm$0.17& 105.1& 194.6& 2.0& 2375.3& 40.5 \\
052948.84-692318.7& 421.7&  3872& 6177.0$\pm$1.1& 246.9$\pm$0.3& 36.4$\pm$ 1.5& 264.6$\pm$ 5.4& 0.126$\pm$0.011& 4.729$\pm$0.069& 0.718$\pm$0.067&  3.921$\pm$0.111& 0.82$\pm$0.06& 0.59$\pm$0.05&  89.1& 107.9& 0.6& 1414.4&  0.1 \\
053141.27-700647.1& 384.1&  3751& 5852.1$\pm$0.8& 215.1$\pm$0.5& 55.7$\pm$ 2.6& 311.0$\pm$ 6.3& 0.054$\pm$0.006& 1.032$\pm$0.128& 0.540$\pm$0.044&  3.458$\pm$0.090& 1.78$\pm$0.12& 0.96$\pm$0.08& 111.1& 134.7& 0.9& 1783.5&  0.9 \\
053156.08-693123.0& 412.2&  3916& 5959.7$\pm$0.9& 252.8$\pm$0.3& 42.2$\pm$ 2.0& 303.6$\pm$ 4.8& 0.069$\pm$0.008& 2.972$\pm$0.098& 0.821$\pm$0.071&  4.149$\pm$0.110& 1.21$\pm$0.07& 1.00$\pm$0.07&  95.5& 120.2& 1.1& 1733.6&  1.7 \\
053202.44-693209.2& 426.5&  4042& 5792.9$\pm$1.7& 234.6$\pm$1.2& 90.0~~~~~~~~~& 285.7$\pm$12.1& 0.078$\pm$0.012& 7.142$\pm$0.158& 0.595$\pm$0.075&  5.279$\pm$0.274& 1.08$\pm$0.15& 0.64$\pm$0.10&  62.0& 121.3& 0.6&  870.0&  0.1 \\ 
053226.48-700604.7& 454.0&  3817& 5899.7$\pm$1.5& 276.2$\pm$0.4& 79.4$\pm$6.1 & 365.8$\pm$ 5.9& 0.221$\pm$0.012& 5.556$\pm$0.058& 0.862$\pm$0.053&  5.139$\pm$0.106& 1.71$\pm$0.10& 1.48$\pm$0.09&  93.3& 143.3& 1.8& 1410.7&  8.7 \\
054006.47-702820.4& 361.5&  3906& 5719.3$\pm$0.8& 286.5$\pm$0.3& 75.4$\pm$14.4& 274.7$\pm$ 2.9& 0.172$\pm$0.006& 4.270$\pm$0.046& 0.630$\pm$0.077&  4.544$\pm$0.099& 1.31$\pm$0.07& 0.82$\pm$0.07&  74.3& 115.2& 0.8& 1032.0&  0.5 \\
\hline
\multicolumn{18}{c}{\textbf{Eccentric orbits~~(\citet{2012MNRAS.421.2616N})} } \\
\hline
052013.51-692253.2& 452.5&  4027& 388.4$\pm$1.5 & 248.0$\pm$1.5& 60.2$\pm$ 1.4& 339.8$\pm$12.6& 0.412$\pm$0.006& 2.846$\pm$0.032& 0.719$\pm$0.095&  6.076$\pm$0.333& 1.50$\pm$0.19& 1.08$\pm$0.15&  73.2& 138.6& 1.1& 1189.1&  2.0  \\
052438.40-700028.8& 411.0&  4125& 321.5$\pm$0.6 & 271.4$\pm$0.3& 90.0~~~~~~~~~& 527.3$\pm$ 3.7& 0.117$\pm$0.005& 4.284$\pm$0.039& 0.675$\pm$0.012&  4.658$\pm$0.037& 6.96$\pm$0.16& 4.70$\pm$0.11& 137.9& 218.0& 2.8& 4787.1& 518.2 \\
052812.41-693417.9& 258.7&  3936& 638.2$\pm$0.7 & 254.1$\pm$1.1& 90.0~~~~~~~~~& 234.6$\pm$ 8.1& 0.236$\pm$0.007& 3.855$\pm$0.043& 0.578$\pm$0.064&  4.807$\pm$0.219& 1.64$\pm$0.18& 0.95$\pm$0.12&  59.4& 100.2& 0.9&  688.7&  0.9  \\
052850.12-701211.2& 662.2&  3796& -52.8$\pm$0.9 & 244.4$\pm$0.1& 51.9$\pm$ 0.9& 581.4$\pm$ 2.8& 0.256$\pm$0.005& 3.436$\pm$0.021& 0.894$\pm$0.023&  4.988$\pm$0.045& 3.18$\pm$0.06& 2.84$\pm$0.06& 160.9& 225.9& 2.7& 4073.4& 77.5  \\
053033.55-701742.0& 390.2&  4081& 800.6$\pm$0.3 & 250.4$\pm$0.1& 54.4$\pm$ 2.2& 480.6$\pm$ 2.5& 0.207$\pm$0.004& 5.090$\pm$0.019& 0.593$\pm$0.028&  4.612$\pm$0.038& 6.15$\pm$0.15& 3.65$\pm$0.12& 127.5& 204.2& 2.3& 3872.6& 200.8  \\
053124.49-701927.4& 541.3&  3848& 420.8$\pm$1.8 & 244.0$\pm$0.2& 90.0~~~~~~~~~& 416.0$\pm$ 4.6& 0.368$\pm$0.009& 3.338$\pm$0.038& 0.464$\pm$0.011&  5.350$\pm$0.058& 2.25$\pm$0.08& 1.05$\pm$0.04&  92.9& 185.8& 1.0& 1473.2&  1.4  \\
053159.96-693439.5& 501.1&  3969& 888.3$\pm$1.1 & 242.3$\pm$0.2& 90.0~~~~~~~~~& 480.7$\pm$ 3.8& 0.388$\pm$0.006& 4.509$\pm$0.025& 0.369$\pm$0.008&  4.918$\pm$0.041& 4.34$\pm$0.10& 1.60$\pm$0.05& 116.4& 224.5& 1.3& 2774.7&  8.8  \\

\end{longtable*}
\end{center}

After finding the best solution according to the procedure above, we
further refined the solution by correcting the input parameters of the
secondary star.  The initial values of these parameters were roughly
estimated using Equations~\ref{kep_eqn} to \ref{eq:omega2},
which include the assumption that the secondary star was in the
middle of main sequence evolution for a star of its mass.  After the
derivation of the orbital solution with the initial parameter values,
the evolutionary status of the secondary was improved by utilizing the
fact that both of the components were born at the same time and have
the same age.  Using the evolutionary tracks of
\citet{2008A&A...484..815B} and the values of $m_1$,~$L_1$ and 
$T_{\rm eff1}$ obtained from the WD code, we estimated the age of the red
giant and this age was used to determine $L_2$ and $T_{\rm eff2}$
since we know $m_2$ from the first orbital solution.  
We then estimated the contribution of each stellar component to the total 
$I$ flux of the system.  In all cases, the secondary contribution 
to the total $I$ flux was less than 4\% and mostly less than 1\%,
meaning that the amplitude of the ellipsoidal light variations in $I$
is not altered by an amount detectable with the current observations
of the $I$ light curve.  However, in a small
number of stars, the depths of
eclipses could be affected at a detectable level ($\ga$0.003 mag.).  
In these cases, the orbital solutions were
re-computed using the values of $T_{\rm eff2}$ and $\Omega_2$
obtained from the parameters of the first solution.
We also checked whether the light from the secondary could affect
our derivation of $T_{\rm eff1}$ from $I$-$K$, which was made
under the assumption that all the $I$ and $K$ flux from the system came from
the red giant primary.  In one system (OGLE 052913.51 -695632.3),
the secondary causes a change of $\sim$50\,K in $T_{\rm eff1}$ while the change
for all other objects is $\la$20\,K and mostly $\la$10\,K.  Since these
changes are less than the estimated error in $T_{\rm eff1}$ of 100\,K
(see \autoref{ss_effects}), we did not make any changes to $T_{\rm eff1}$.

\begin{figure*}
\begin{center}
        \includegraphics[scale=0.42]{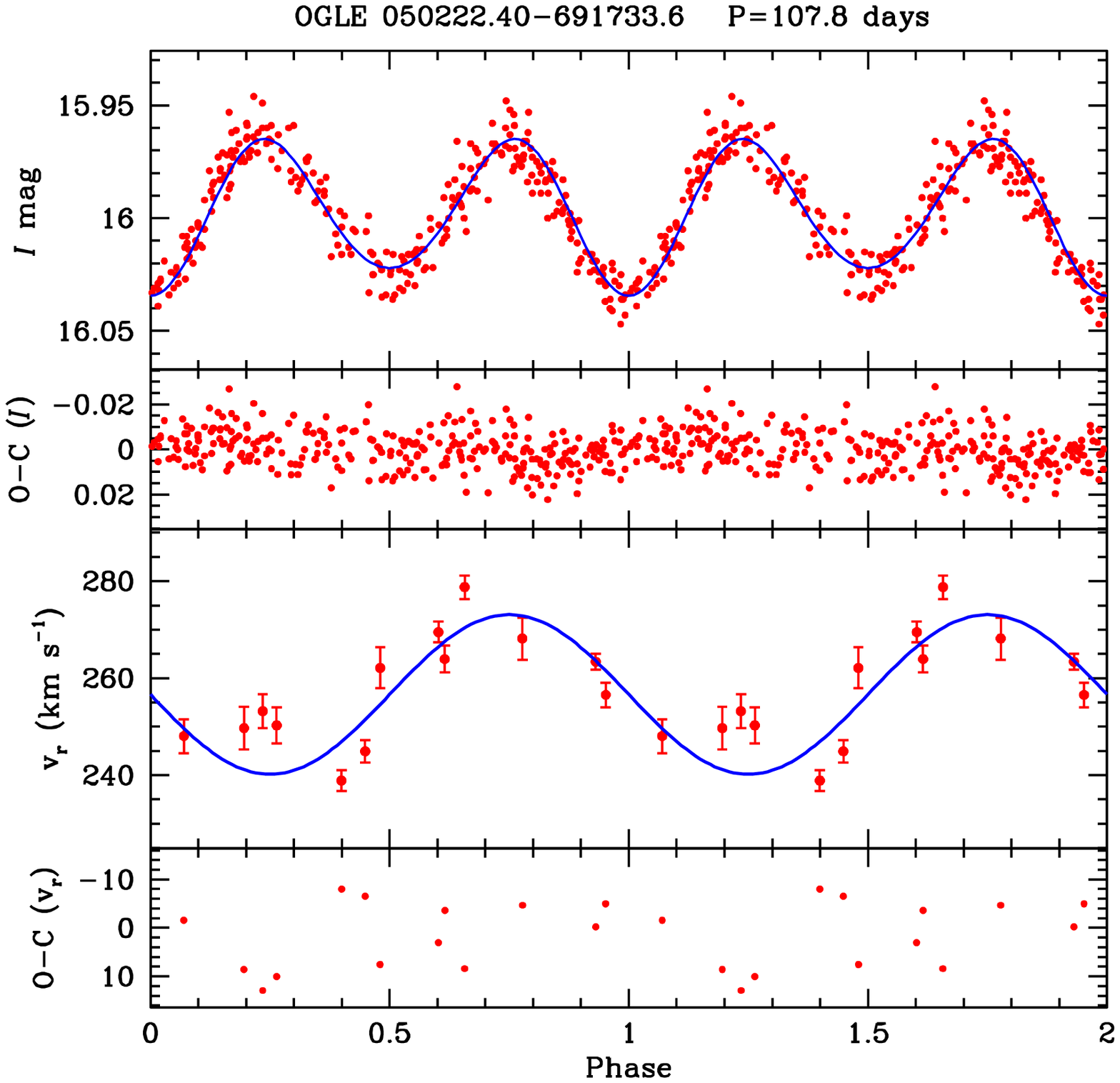}
	\includegraphics[scale=0.42]{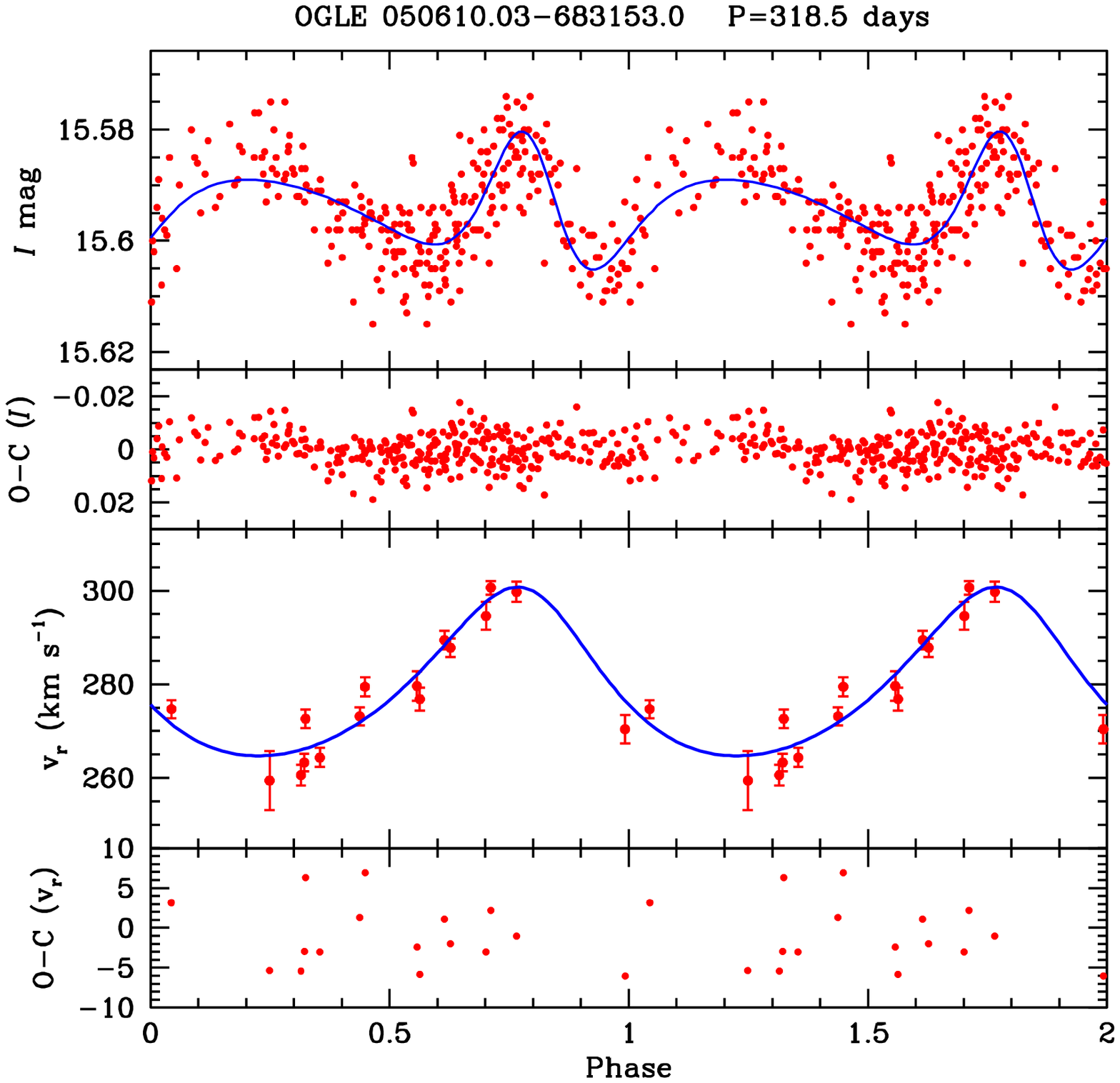}
	\caption{ Fits of the light and velocity curves with the best solution. 
In the top and third panels, red points are 
the observations and blue curves are the fits for light and velocity curves, 
respectively. In the second and bottom panels, red points are the fitting 
residual for light and velocity curves. The 
two examples are the first object of \autoref{tb:best} with circular and eccentric orbits, 
respectively. The fittings of the whole sample are provided online.
\label{fig:fitting}}
\end{center}
\end{figure*}

We attempted to model 97 objects in total but for various reasons 16 objects 
were rejected. 
Detailed remarks noting the reasons for rejection are given in \autoref{reject}. 
The parameters of 
the solutions for the successfully modelled objects (81 in total) are given in \autoref{tb:best}
along with standard errors for the parameters.  We only provide values of $L_2$
and $R_2$ to 0.1 solar units since changes of this size do not
affect the solution at a detectable level and the only parameter
for the secondary star that we derive from the orbital solutions is $m_2$.
The light and velocity curve fits for all the successful solutions are shown 
in \autoref{fig:fitting}. The full version of \autoref{fig:fitting} is provided online.

\subsection{A check on error estimation by the WD code}

\begin{figure*}
\includegraphics[width=15cm]{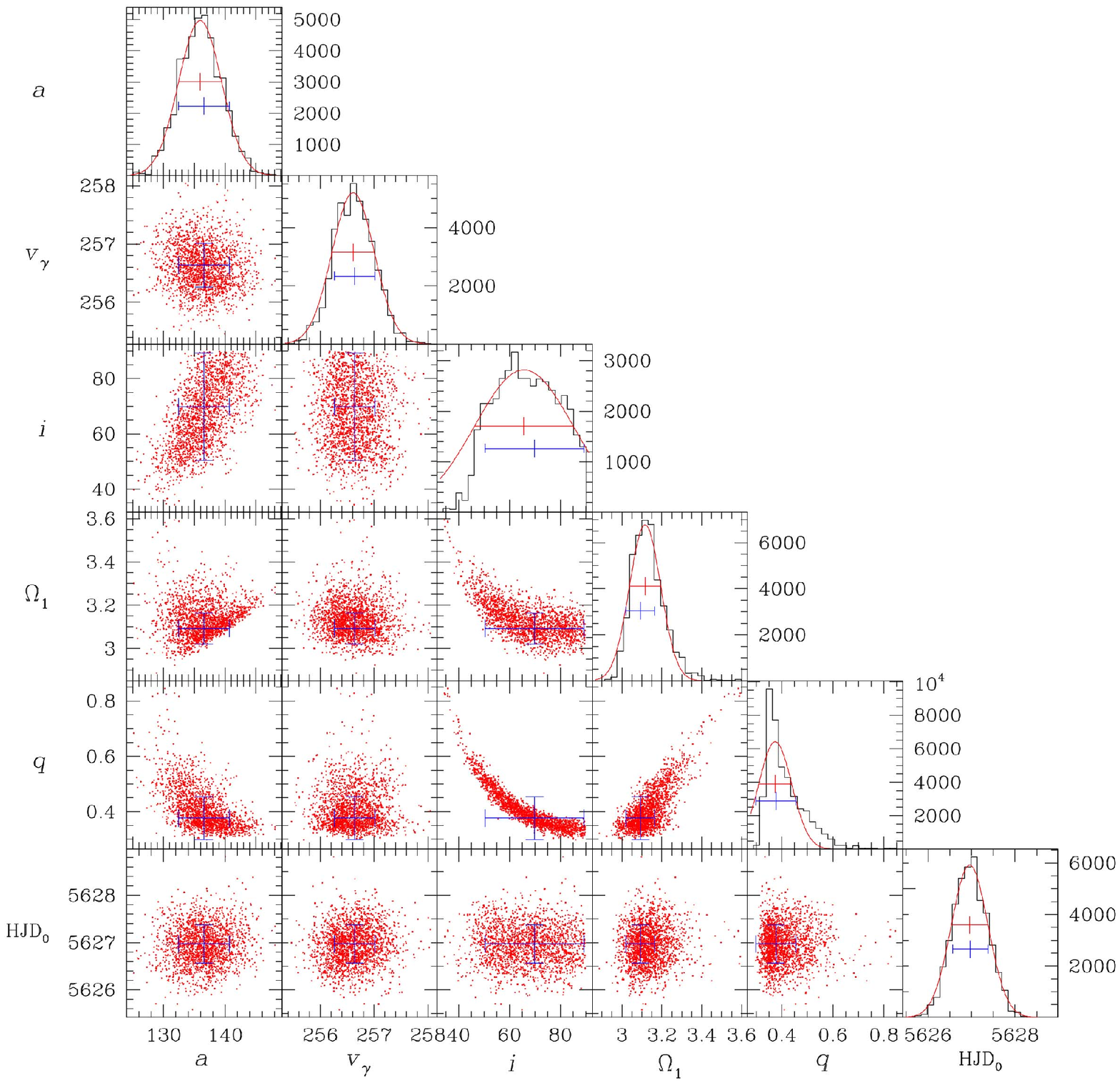}
\caption{The results of MCMC simulation for OGLE 050222.40 -691733.6.  
The scatter plot panels below the diagonal show one red point
for each proposed pair of values of the two parameters pertaining to the panel.
The density of points is proportional to the probability that the parameters
could produce the observed radial velocity and $I$ values.  Clear correlations
between some pairs of parameters are seen.  The panels on the diagonal of the
plot show the marginal distribution (black histogram) of the associated parameter for all
proposals in the MCMC simulation.  A gaussian has been fitted to points in the 
marginal distribution with values more than 25\% of the maximum (red curve) 
and the mean value and 1-$\sigma$ error bars are shown (red line).  
The blue error bars are the 1-$\sigma$ error
estimates given by the WD code.}
\label{fig:mcmc1}
\end{figure*}

\begin{figure*}
\includegraphics[width=15cm]{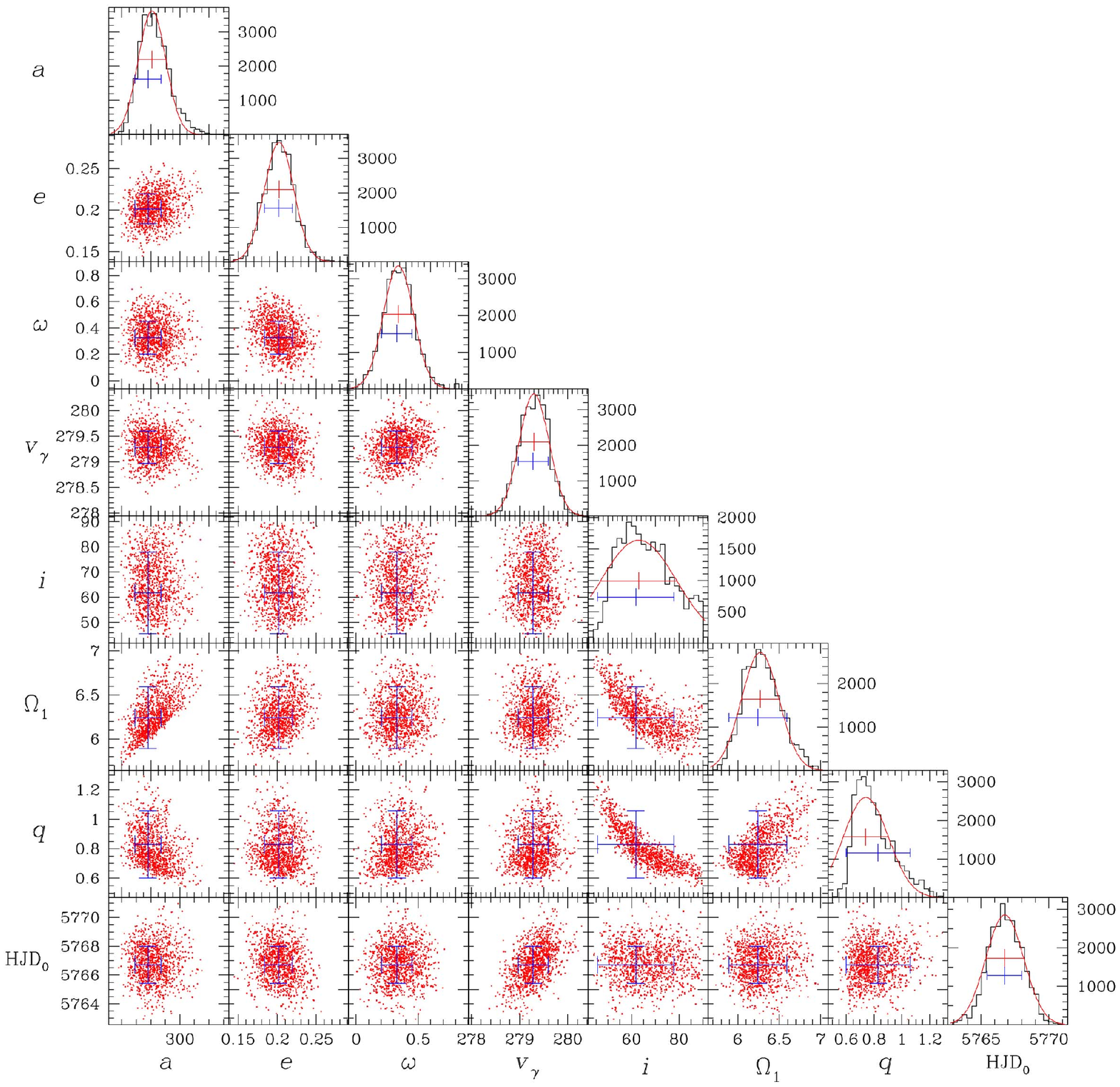}
\caption{The same as \autoref{fig:mcmc1} but for the OGLE 050610.03 -683153.0,
which has an eccentric orbit.}
\label{fig:mcmc2}
\end{figure*}

The referee of this paper expressed doubts about the reliability of the
error estimates given by the WD code.  In order to provide independent
error estimates for the model parameters, we have used an adaptive 
Markov chain Monte Carlo (MCMC) method to examine the distribution
of likely parameter values.

The basic MCMC code we used is that described by \citet{haario2001}
and \citet{haario2006}\footnote{The fortran 90 code is available
at http://helios.fmi.fi/~lainema/mcmcf90/.}.  The user provides the residual
sum of squares function for which we used the value given by the WD code.
The residual sum of squares was normalized to the value for the best WD
model so that the normalized value gives $-2 \ln(p)$, where $p$ is the
likelihood of the model relative to the best model.  For each set of model
parameters proposed by the MCMC code, the WD code was run to obtain the
residual sum of squares and the process was repeated for the defined number
of proposals.  Since tens of thousands of proposals are required to generate
a useful probability density function when six (circular orbits) or
eight (eccentric orbits) parameters vary, the method is quite computationally
expensive.

In Figures~\ref{fig:mcmc1} and~\ref{fig:mcmc2}, we illustrate the results 
of our MCMC simulations for OGLE 050222.40 -691733.6,
which has a circular orbit, and OGLE 050610.03 -683153.0, which has an
eccentric orbit.  We generated 50000 proposals for the first object and
30000 for the second.  In both figures, there are clear correlations between some
parameter, for example $q$ and $i$, as well as some forbidden regions ($i > 90$,
or combinations of $\Omega_1$ and $a$ that overfill the Roche Lobe).
In all cases, the most likely parameter values estimated from the marginal
distributions on the diagonals of the figures agree with the values from
the WD code to less than one standard deviation as estimated by the WD code.
In fact the agreement is to within a small fraction of one standard deviation
except possibly for $q$ where the marginal distribution has a large skewness.  Note
also that the standard deviations estimated by the WD code and the MCMC analysis
agree well, once again with the possible exception of $q$ due to the highly 
skewed distribution resulting from the MCMC analysis.  

We have run the MCMC code for a small number of other objects and find similar
results to those shown for OGLE 050222.40 -691733.6 and OGLE 050610.03 -683153.0 
(to run the MCMC code on all 81 objects would be
too computationally expensive).  Our conclusion is that both the parameter estimates and
the standard errors given by the WD code are reliable.

\subsection{Rejected objects} \label{reject}

We rejected the orbital solutions of 16 objects. 
The reasons for rejection are given below for each of these objects:\\
\textit{OGLE 050107.08 -692036.9} ---
The solution only converged for $i=$ near 90 degrees, with derived 
mass $m_1 \sim 1.9$\msolar~and $q>1$, and for these solutions the 
fitting of the light and radial velocity curves is not good.
Since $T_{\rm eff1} = 5773~{\rm K}$, the primary star is a hot star. A star of this 
temperature should be in the pre-RGB phase. It is unlikely to be in the He core burning 
phase, as it would have been larger earlier in its life. It could also be a post-AGB star 
in principle, but this is unlikely because $m_1$ is much larger than the post-AGB star
mass ($\sim$0.55\msolar) for the observed luminosity of 3278\lsolar.  If it is on the pre-RGB, 
it needs $m_1$$\sim$6.5\msolar~to match the observed luminosity.
However, the solution has $m_1$$<$2.0\msolar. \\
\textit{OGLE 050334.97 -685920.5} ---
From the relatively hot effective temperature (5666\,K) 
and the observed luminosity (2526\lsolar),
this star should be  
on the pre-RGB phase with mass $\sim$6.3\msolar.  However, the large velocity 
amplitude and relatively short period require $m_1$$\sim$2.6\msolar~and 
a large mass for $m_2$ ($\sim$7\msolar), so that $q>1$.
We are unable to get an acceptable solution for this object. \\
\textit{OGLE 050454.49 -690401.2} ---
This is a hot star and the effective temperature (6017\,K) and 
luminosity (5175\lsolar) are consistent with a pre-RGB star of mass 7.0\msolar. 
However, the large velocity
amplitude and long period require a large mass for $m_2$ ($\ga$10\msolar) so that $q>1$.
We are unable to get an acceptable solution for this object.\\
\textit{OGLE 050758.17 -685856.3} ---
The solution with $i\sim75$ degrees has the lowest light curve dispersion, but the calculated 
$m_1$ (7.88\msolar) is too large since even at the base of the giant branch the 
evolutionary track for this $m_1$ gives a luminosity that is larger than the observed 
luminosity. For the observed luminosity of this object, evolutionary tracks require the 
mass of $m_1$ to be less than 7.0\msolar. \\
\textit{OGLE 050800.52 -685800.8} ---
This is similar to the previous object. The mass of $m_1$ should be less than 
8.0\msolar, but the derived solution shows $i\sim71$ and $m_1$=9.12\msolar. \\ 
\textit{OGLE 051515.95 -685958.1} ---
We were not able to get a solution to fit the two very different light maxima. \\
\textit{OGLE 051845.02 -691610.5} ---
All solutions have $q \gg 1$ and $R_1 \ll R_{L,1}$ (meaning no stable mass transfer is
possible), with the best solution having $i \sim 60$ degrees, 
$m_1 \sim 0.8$\msolar~and $m_2 \sim 1.9$\msolar.\\
\textit{OGLE 052032.29 -694224.2} --- 
This is a hot star with $T_{\rm eff1} = 5972$ K, suggesting the star is in the pre-RGB 
phase. Given the luminosity of 2569\lsolar, $m_1$ should then be $\sim$6\msolar.
The orbital solutions have about half this mass for $m_1$.  We are unable to find an
acceptable solution for this object.\\
\textit{OGLE 052048.62 -704423.5} ---
From the light curve and low amplitude velocity curve, 
this appears to be a pulsator of very low amplitude, 
not an ellipsoidal binary. \\
\textit{OGLE 052203.16 -704507.8} ---
The orbital solution requires $m_1 < 0.4$\msolar, which is too low a mass for the star
to have evolved off the main sequence. In addition, the observed velocity curve shows 
inconsistent points (sharp dip, low amplitude) and the scatter of the light curve 
suggests pulsation. This is possibly a sequence D star
\citep{1999IAUS..191..151W,2010MNRAS.405.1770N}. \\
\textit{OGLE 052228.84-694313.8} ---
We were not able to get an acceptable solution to fit the two light maxima
of different brightness.  This is the same problem that we encountered for
OGLE 051515.95 -685958.1. \\
\textit{OGLE 052458.88 -695107.0} ---
The light curve suggests pulsation and the velocity amplitude is about 5 km s$^{-1}$, 
only a little larger than  the mean error of all the data
(2 km s$^{-1}$).  Once again, this is probably a sequence D star. \\
\textit{OGLE 052513.34 -693025.2} --- This is a very small amplitude 
object so the photometry errors are large compared to the amplitude.  It is 
impossible to put any meaningful constraint on the inclination.\\
\textit{OGLE 052757.39 -693901.7} ---
This is an object from \cite{2010MNRAS.405.1770N}. The light curve amplitude is large, 
up to 0.6 mag in the $I$ band, very different from all the other ellipsoidal red giant 
binaries in this paper.  The system seems to be an eclipsing pair of red giants.  
The strategy described 
in \autoref{method} for orbit solutions is not applicable to this binary system.  Owing to 
this, we set the gravity darkening exponent $g_2=0.35$ and the bolometric albedo A$_2$=0.5
(appropriate for a red giant secondary) and adjusted $T_{\rm eff1}$, $T_{\rm eff2}$, 
$\Omega_1$ and $\Omega_2$ to get a solution.
The solution only converged at $i=90$ and 80 (with strong eclipses), with very good 
fitting. The secondary red giant has a mass similar to that of the primary red giant, but 
it has a smaller size and fainter luminosity. Due to the somewhat arbitrary selection
of input parameters and the unusual nature of this object, we reject the solution.\\
\textit{OGLE 052928.90 -701244.2} ---
The light curve of this star shows a large pulsation amplitude along with the binary 
ellipsoidal light variations. In addition, the data quality of the whole light curve 
is poor, so it is hard to determine the orbital period. Its periods from OGLE II data 
and MACHO data are different. When using the period determined from OGLE II data, 
$i=50$ gives the best solution. However, then the mass $m_1$$\sim$0.35\msolar, too 
low for a red giant. \\
\textit{OGLE 053219.66 -695805.0} ---
From the OGLE II finding chart, there are two stars 1.3\arcsec apart, too close for the 
WiFeS (Wide Field Spectrograph) instrument \citep{2007Ap&SS.310..255D,2010Ap&SS.327..245D} 
to resolve consistently, so the radial velocity curve might be a combination of these 
two stars. \\

\subsection{The effect of $T_{\rm eff}$, gravity brightening, 
limb darkening and distance modulus on orbital parameters}\label{ss_effects}

\tabletypesize{\small}
\begin{table*}
\centering
\caption{The effect of $T_{\rm eff}$, $g_1$, limb darkening and distance modulus on orbital parameters of OGLE 050222.40 -691733.6}
\begin{tabular}{ccccccccccc}
\tableline
\tableline
Model & DM & $T_{\rm eff}$ & $R_1$ & $g_1$ & LD & $a$ & $q$ & $i$ & $m_1$ & $m_2$ \\
& & K & ${\rm R_{\odot}}$ & & & ${\rm R_{\odot}}$ & & $\deg$ & ${\rm M_{\odot}}$ & ${\rm M_{\odot}}$ \\
\tableline
 1 & 18.49 & 3990 & 51.59 & 0.301 & -3 & 136.6$\pm$4.1 & 0.3765$\pm$0.0781 & 69.8$\pm$19.5 & 2.139$\pm$0.230 & 0.805$\pm$0.188 \\
 2 & 18.49 & 4090 & 48.28 & 0.301 & -3 & 129.8$\pm$3.8 & 0.4027$\pm$0.0826 & 70.2$\pm$19.3 & 1.803$\pm$0.192 & 0.726$\pm$0.168 \\
 3 & 18.49 & 3990 & 51.53 & 0.327 & -3 & 136.8$\pm$4.6 & 0.3847$\pm$0.1048 & 67.3$\pm$22.9 & 2.136$\pm$0.271 & 0.822$\pm$0.247 \\
 4 & 18.49 & 4090 & 48.28 & 0.327 & -3 & 130.4$\pm$4.0 & 0.3998$\pm$0.0841 & 70.3$\pm$19.9 & 1.832$\pm$0.200 & 0.732$\pm$0.174 \\
 5 & 18.49 & 3990 & 51.60 & 0.301 & -2 & 136.7$\pm$4.1 & 0.3762$\pm$0.0779 & 69.8$\pm$19.4 & 2.143$\pm$0.229 & 0.806$\pm$0.188 \\
 6 & 18.49 & 3990 & 51.59 & 0.301 & -1 & 136.3$\pm$4.1 & 0.3778$\pm$0.0777 & 69.8$\pm$19.2 & 2.124$\pm$0.227 & 0.802$\pm$0.186 \\
 7 & 18.54 & 3990 & 52.72 & 0.301 & -3 & 138.4$\pm$4.6 & 0.3795$\pm$0.1011 & 67.1$\pm$22.2 & 2.220$\pm$0.276 & 0.842$\pm$0.248 \\
\tableline
\end{tabular}
\label {tab_params_t1_gr1_dm}
\end{table*}

In the modelling process, the red giant size is determined
by the assigned values of $T_{\rm eff1}$ and $L_{\rm 1}$.
For an observed amplitude of
ellipsoidal light variability, the assigned size of the 
red giant will
clearly affect the derived orbital separation and this will in turn affect the
derived masses for the given orbital period.  Here, we examine
the effect of $T_{\rm eff1}$ and adopted distance modulus (DM), which affects $L_{\rm 1}$,  on
derived orbital parameters.  Also, since the relative depths of the alternating
minima depend on the amount of gravity brightening and the limb 
darkening (see \autoref{best_sol}), 
we examined the effect of the gravity brightening exponent $g_1$ and the limb
darkening law (LD) on the orbital solutions.
We illustrate the effects using the system OGLE 050222.40 -691733.6 and the
results are shown in \autoref{tab_params_t1_gr1_dm}.  For each set of
input values of $T_{\rm eff1}$, $g_1$, LD and DM, the derived orbital parameters
are given along with the standard error $\sigma$ given by the WD code.

Model 1 is the reference model obtained using standard parameters.
Model 2 shows the effect of increasing the assigned value of $T_{\rm
  eff1}$ by 100\,K.  We note that $T_{\rm eff1}$ is derived from the
$I$-$K$ color and the relation between $I$-$K$ and $T_{\rm eff}$ given
in \citet{2000AJ....119.1424H} and \citet{2000AJ....119.1448H}.  A
change of 100\,K for $T_{\rm eff}$ values around $T_{\rm eff} \sim
4000$\,K corresponds to a change in $I$-$K$ of $\sim$0.1.  The
observed average value of $I$ obtained from the OGLE light curve is
very accurate with an error less than 0.01 mag.  The $K$ value from
2MASS was taken at a random orbital phase so that the measured
value could differ from the mean value by up to half the $K$ amplitude.  
Since the median $K$ amplitude of our objects is 0.048 mag (we assume 
the $K$ amplitude is equal to the $I$ amplitude, which is readily
obtained from the $I$ light curve), the maximum difference between the observed and
average values of $K$ should be typically $0.5*0.048=0.024$ mag.  The photometric
errors in $K$ range from $\sim0.02-0.09$ mag for the brightest to the
faintest objects.  Thus the maximum expected total error in $K$ and
$I$-$K$ is $\sim$0.1 mag, the latter corresponding to a maximum error
in $T_{\rm eff1}$ of $\sim$100\,K.

It can be seen from \autoref{tab_params_t1_gr1_dm}  
that increasing $T_{\rm eff1}$ by 100\,K decreases the size of the
red giant and the orbital separation by $\sim$5\%, the mass $m_1$ by
$\sim$16\% ($\sim$1.6 $\sigma$) and the mass $m_2$ by 
$\sim$10\% ($\sim$0.4 $\sigma$).

As shown in \citet{1997A&A...326..257A}, the gravity brightening
exponent $g_1$ varies with $T_{\rm eff}$.  In model 3, we examine the
effect of changing $g_1$ by 0.026, which corresponds to the change in
$g_1$ for an increase in $T_{\rm eff}$ of 100\,K, and which is also
similar to the scatter in the derived values of $g_1$ at a given
temperature around 4000\,K in the models of
\citet{1997A&A...326..257A}.  The results in
\autoref{tab_params_t1_gr1_dm} show that changes in $g_1$ of this size
have almost no effect on the derived orbital parameters.  Since
$T_{\rm eff}$ and $g_1$ are correlated, in model 4 we show the effect
of the combined change in $T_{\rm eff}$ and $g_1$.  It can be seen
that the solution parameters are close to those resulting from the
change in $T_{\rm eff}$ alone.

Models 1, 5 and 6 show the effects that changes in the limb darkening 
law make to the derived orbital parameters.  Model 1 uses our
standard assumption of a square root law (LD = -3 in the WD code),
model 5 uses a logarithmic law (LD = -2) and model 6 used a
linear cosine law (LD = -1).  It can be seen that the adopted limb
darkening law makes almost no difference to the solution
parameters.

Finally, we examined the uncertainties caused by the uncertainty in
the LMC distance modulus.  Note that in this paper we use a distance
modulus of 18.49 \citep{2014AJ....147..122D} whereas in our data paper
\citep{2014AJ....148..118N} we used a distance modulus of 18.54
\citep{2006ApJ...642..834K}.  Model 7 in
\autoref{tab_params_t1_gr1_dm} shows the effect of increasing the
distance modulus by 0.05.  This change leads directly to a slightly
larger luminosity and radius for the red giant and subsequently to
slightly larger values for $a$, $m_1$ and $m2$.  The masses $m_1$ and
$m2$ increase by $\sim4$\%, well with the 1-$\sigma$ standard errors for
these quantities.

In summary, it seems that the uncertainties in $T_{\rm eff}$ have the
greatest influence on the orbital solution parameters.  The estimated
maximum uncertainty in $T_{\rm eff1}$ of 100\,K could lead to
uncertainties in $m_1$ and $m_2$ of $\sim$16\% and $\sim$10\%,
respectively, corresponding to $\sim$1.6 $\sigma$ and 0.4 $\sigma$.
Note that the errors given in \autoref{tb:best} are model fit 
errors only and they do not include the additional uncertainties
arising from uncertainties in $T_{\rm eff}$, gravity darkening, limb 
darkening and distance modulus, which have been discussed here.

\section{Analysis of results}

\autoref{tb:best} provides orbital solutions
for a large number (81) of ellipsoidal red giant binaries in the LMC, which 
we refer to hereinafter as our sample.  
We now analyse the properties of our sample.

\subsection[]{The evolutionary status of the red giant primaries}\label{s:evolution_status}

In general, our sample consists of binaries of two types: (1) those containing
low mass red giants ($m_1 \leq 1.85$\msolar) 
which are evolving on the RGB or AGB with degenerate He or C/O cores, respectively, 
and (2) those containing intermediate mass red giants (1.85\msolar~$ \leq m_1 \leq 8.0$\msolar) 
evolving with non-degenerate He cores prior to core He burning or evolving on the
AGB after core He burning (since red giants shrink after He ignition in the
core, it is unlikely that they would be ellipsoidal variables during the
core He burning phase).  

\begin{figure}
\begin{center}
\includegraphics[width={1.0\columnwidth}]{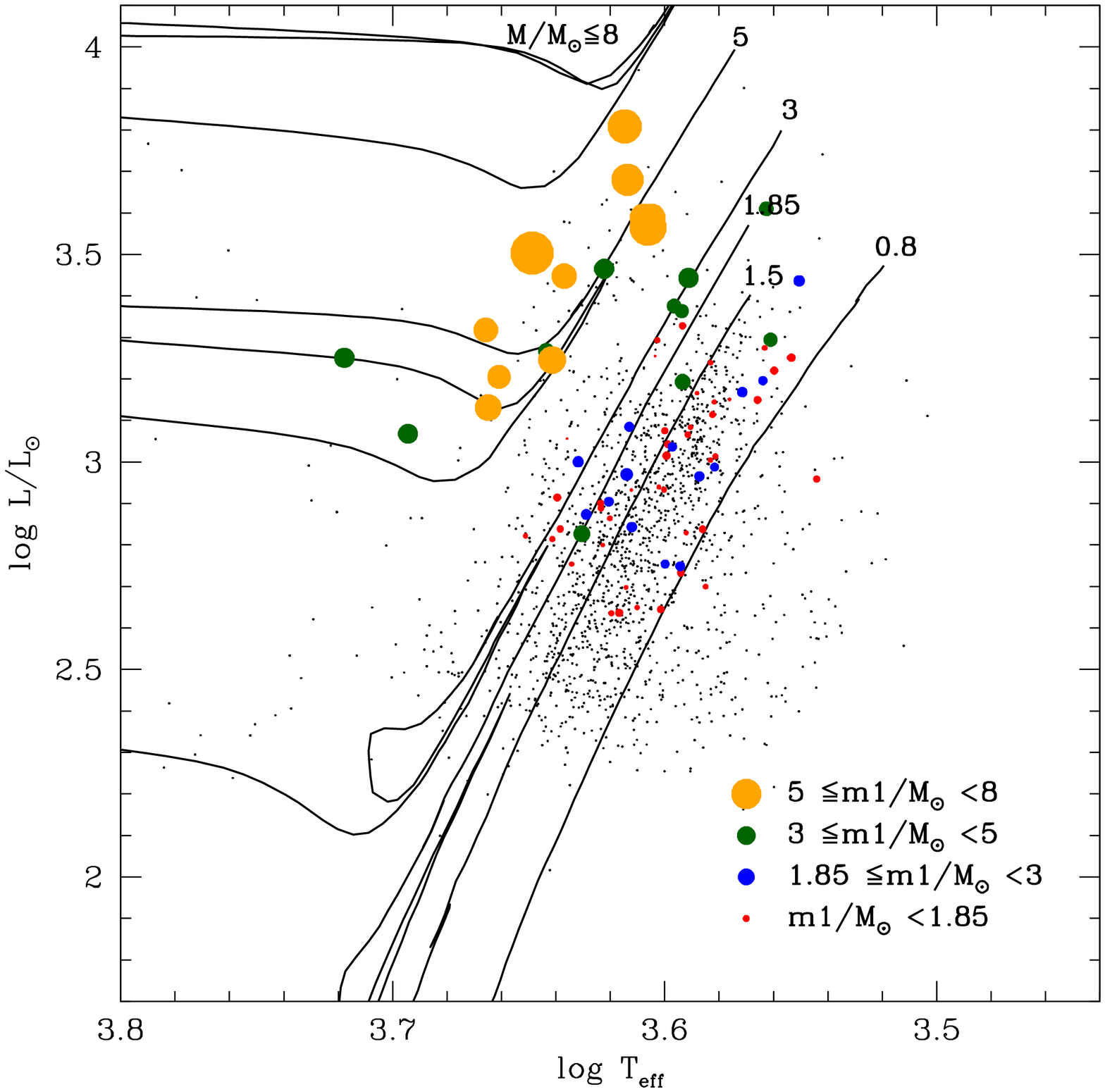}
\caption{
The HR diagram for red giant ellipsoidal variables.  Ellipsoidal
variables belonging to the full sample in \citet{2004AcA....54..347S}
are shown as black dots, with $T_{\rm eff}$ and $\log
L$/\lsolar~obtained from $I$ and $K$ photometry as described in
\citet{2012MNRAS.423.2764N}.  The objects in our sample are shown by
symbols whose color and size depends on the red giant mass.  The
symbol diameter is proportional to mass $m_1$.  The evolutionary
tracks of \citet{2008A&A...484..815B} are shown as black lines.  For
the 0.8 and 1.5\,\msolar tracks, the evolution is shown only to the
RGB tip while for the more massive tracks AGB evolution to near the
first thermal pulse is included as well.
}
\label{hrd}
\end{center}
\end{figure}

\autoref{hrd} shows our red giant
binaries in the HR diagram with the masses indicated by symbols of different
size and color.  Evolutionary tracks for stars of selected masses are also shown.
Here, a metallicity of $Z=0.008$ is used for the tracks of the younger stars 
with $M \ge 3$\,\msolar~ while a metallicity of $Z=0.004$ is used for the 
older stars of $M$=0.8, 1.5 and 1.85\,\msolar ~in order to take account of the
age-metallicity relation for LMC stars \citep{2013AJ....145...17P}.
In general, the masses derived for the red giants fall in the positions
expected from evolutionary tracks, although there will be some
scatter caused by errors in the determined mass and by metallicity variations.
We note that these results, together
with recent asteroseismological results from Kepler and CoRoT 
\citep[e.g.][]{2010A&A...522A...1K,2013ApJ...765L..41S}, provide rare tests
for the positions of red giants with a wide range of masses in the HR diagram.

\begin{figure}
\begin{center}
        \includegraphics[width={1.0\columnwidth}]{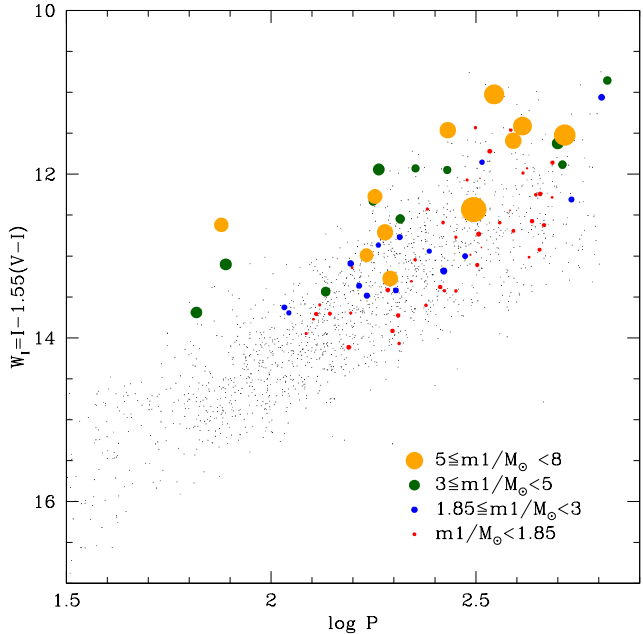}
\caption{The Period-Luminosity diagram for ellipsoidal
	variables showing how position depends on the red giant mass $m_1$. 
        $W_I$ is a reddening-free Weisenheit index \citep{1982ApJ...253..575M}
        that is essentially a measure of luminosity
        and P is the orbital period in days.
	Symbols are the same as in \autoref{hrd}.}
\label{period-lum}
\end{center}
\end{figure}

\begin{figure}
\begin{center}
        \includegraphics[width={1.0\columnwidth}]{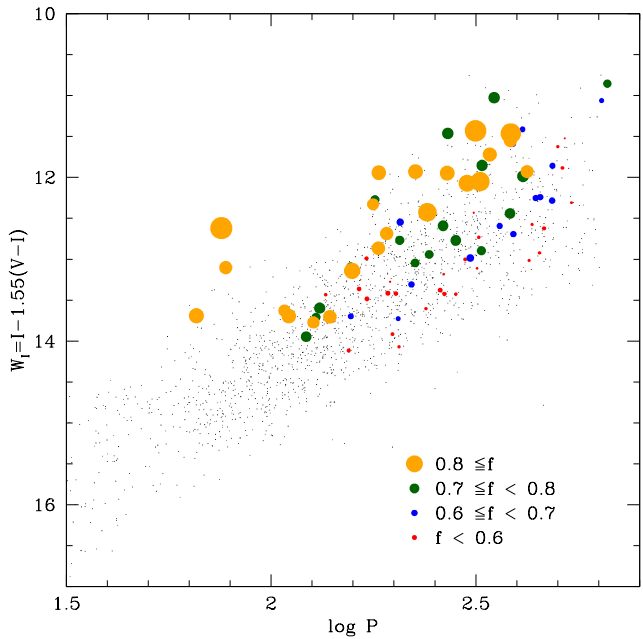}
\caption{The Period-Luminosity diagram similar to \autoref{period-lum} except that the
symbol sizes and colors now indicate the Roche lobe filling factor $f$,
with symbol diameter proportional to $f^3$. }
\label{pl-f}
\end{center}
\end{figure}

In \autoref{period-lum}, we show our sample
of ellipsoidal variables in the Period-Luminosity diagram, with the
masses $m_1$ for each system indicated by the symbol color and size.
The red giant mass generally increases as the luminosity increases at
a given period.  This verifies the suggestions of
\citet{2004AcA....54..347S} and \citet{2014AJ....148..118N} that,
because higher mass systems will have larger separations at a given
orbital period, the red giant primaries in them will need to get to a 
higher luminosity (radius) to produce detectable ellipsoidal light variations.  
Note that our sample is somewhat biased to higher masses as
\citet{2014AJ....148..118N} deliberately chose extra objects on the
high-luminosity side of the PL sequence in order to get a reasonable
number of intermediate mass stars in the sample.

Metallicity should also affect the position of a
binary system on the PL sequence.  If we consider two systems with
stars of the same mass but different metallicity, then the orbital
periods and separations will be the same but the one with lower
metallicity will have a smaller red giant at a given luminosity, so the
red giant will need to get to a higher luminosity to reach the radius
required to produce detectable ellipsoidal light variations.  This
will lead to relatively lower masses on the higher luminosity side
of the PL sequence for lower metallicity stars.  Examples of such
stars can be seen in \autoref{period-lum}, especially for the systems
with $m_1 < 1.85$\msolar. 

Finally, we note that in addition to systematic variations in mass at 
a given period in the PL diagram, a mass dispersion is also
expected.  This is because ellipsoidal variations can be detected over
a range of values for the Roche lobe filling factor.  Hence, for an individual system
with a given orbital period, ellipsoidal variations will be
detectable over a range in luminosities.  The ellipsoidal light amplitude should
increase with luminosity in this system.  This effect was confirmed by
\citet{2004AcA....54..347S} who demonstrated that higher
amplitudes of ellipsoidal light variations were shown by stars on the
higher luminosity side of the PL sequence.  The origin of the higher light
amplitude is a larger Roche lobe filling factor $f$.  Since we determine $f$ for all
the stars in our sample we can test this assumption directly.  \autoref{pl-f}
shows the PL diagram with the value of $f$ for our sample indicated by the symbol
size and color.  There is a clear increase in $f$ to the high luminosity side
of the PL sequence, as expected.  By comparing \autoref{period-lum} and \autoref{pl-f}
it can be seem that the systems with the larger values of
$f$ are not necessarily the ones with higher mass.

\subsection{The primary mass distribution}\label{s:m1-dist}

\begin{figure}
\begin{center}
\includegraphics[width={1.0\columnwidth}]{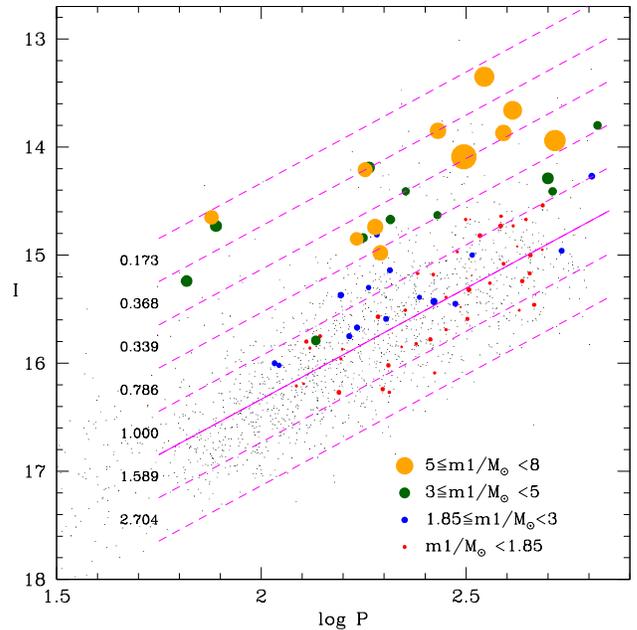}
\caption{The $I$-$\log P$ diagram with symbols as in \autoref{hrd}.  The solid pink line
is a fit in the interval $1.75 < \log(P) < 2.85$ to the sample of ellipsoidal variables in
\citet{2004AcA....54..347S} (black dots). The dashed lines are spaced from the
solid line at intervals of 0.4 magnitudes in $I$ and the numbers in the left
part of the plot are the weights applying to observed systems lying between
the adjacent pair of lines.}
\label{i-logp+m1}
\end{center}
\end{figure}

\begin{figure}
\begin{center}
\includegraphics[width={1.0\columnwidth}]{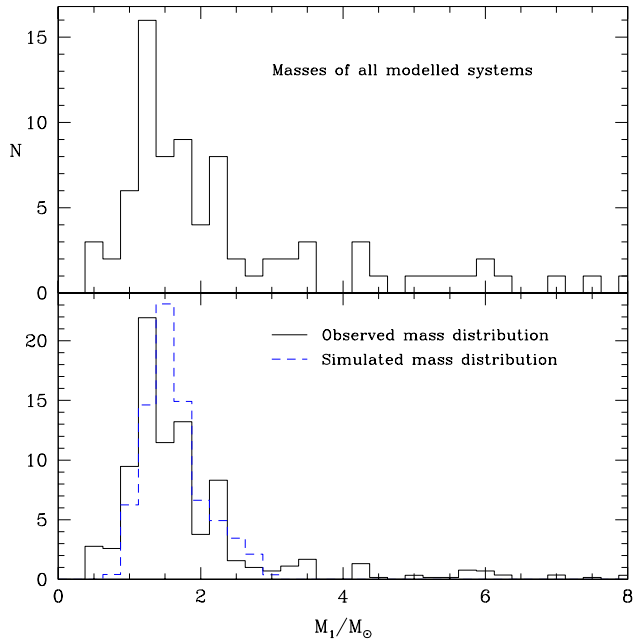}
\caption{Distribution of the primary mass. The top panel shows the raw distribution
for all red giant masses in our sample while the bottom panel shows the distribution 
adjusted for selection bias as described in the text.  Also shown in the bottom panel 
is the equivalent distribution of red giant masses in sequence E stars according to a 
simulation based on the method of \citet{2012MNRAS.423.2764N}.  The simulated 
distribution is normalized to have the
same number of objects with $m_{\rm 1} < 3$\msolar~ as the observed distribution.
}  
\label{m1dist}
\end{center}
\end{figure}

As noted in \citet{2014AJ....148..118N}, our sample of observed objects was selected
from the $I$-$\log P$ plane.  The objects for which we were able to derive complete 
solutions are shown in this plane in \autoref{i-logp+m1} where the red giant mass $m_{\rm 1}$
is indicated by the symbol color and size.  It is clear that $m_{\rm 1}$ depends
strongly on position in this diagram (as it did in the $W$-$\log P$ plane, 
\autoref{period-lum}).

The top panel of \autoref{m1dist} shows the raw distribution of red
giant masses for all objects in our full sample.  This distribution is
biased by observational selection, especially by the deliberate
selection of brighter objects at a given period by \citet{2014AJ....148..118N}
and by the relatively bright objects from \citet{2012MNRAS.421.2616N}
Our aim is to make an unbiased
estimation of the distribution of primary red giant masses in the full
sample of LMC ellipsoidal variables of \citet{2004AcA....54..347S}.
The full sample of the ellipsoidal variables (the sequence E stars) are shown in
\autoref{i-logp+m1} as small black dots.  The solid pink line is
a linear fit to this sample in the interval $1.75 < \log(P) < 2.85$.

Looking at \autoref{i-logp+m1}, we see that the 
masses $m_{\rm 1}$ of stars in our sample appear to be roughly similar on
lines parallel to the $I$-$\log P$ sequence formed by the black dots
in \autoref{i-logp+m1} i.e. on lines parallel to the solid pink line.
At a given period, the more massive stars are brighter
so that without a correction for the preferential selection
of brighter object, the raw mass distribution will over-emphasize
the number of high mass objects.

In order to attempt to remove our selection bias towards brighter objects
at a given period, we
have divided the $I$-$\log P$ plane into strips parallel to the
$I$-$\log P$ sequence, with the strip width being 0.4 magnitudes in
$I$.  In each of these strips there are a number of objects from our
sample for which we have estimates of $m_{\rm 1}$, as well as many
objects from the sample of \citet{2004AcA....54..347S}.  In order to
estimate the mass distribution in the unbiased sample of
\citet{2004AcA....54..347S} in the interval $1.75 < \log(P) < 2.85$, where
our objects lie, we assign to each of our objects in a
given strip a weight $w_{\rm i}$ which is the ratio of the number of
\citet{2004AcA....54..347S} objects with $1.75 < \log(P) < 2.85$ in the strip to the number objects
from our sample in this strip, scaled by a common factor.  These
weights are shown in \autoref{i-logp+m1} on the left side of each
strip, and they clearly show our preferential selection of
brighter objects.  Applying these weights, we derive a histogram showing our
estimate of the distribution of red giant masses in the sample of
\citet{2004AcA....54..347S} in the interval $1.75 < \log(P) < 2.85$.  
This histogram is shown in the bottom
panel of \autoref{m1dist} where the number of objects assigned to each
bar of the histogram is $\sum_{i}^{} w_{\rm i}$ for all those objects
$i$ whose $m_{\rm 1}$ lies between the mass limits for the bar of the
histogram.

Also shown in the bottom panel of \autoref{m1dist} is the distribution
of red giant masses expected for sequence E stars according to a
simulation based on the modelling prescriptions described in
\citet{2012MNRAS.423.2764N}.  We have assumed that the comparison
sample of sequence E stars from the simulation consists of all
observable sequence E stars brighter than $M_{\rm bol} = -1.6$ 
(i.e. two magnitudes below the tip
of the RGB  at $M_{\rm bol} = -3.6$) but excluding low mass stars ($m_{\rm 1} < 1.85$\msolar)
that are on the AGB and which have become more luminous than the RGB
tip.  These stars were excluded because they
are observed to be pulsationally variable
\citep[e.g.][]{2015MNRAS.448.3829W} and their ellipsoidal variability
will be difficult to detect when combined with semiregular pulsational
variability. 

Comparing the raw mass distribution of the red giants (top panel of
\autoref{m1dist}) with the estimated true distribution (bottom panel),
it can be seen that the raw observed sample is biased towards the
brighter, more massive stars.  This bias was described by
\citet{2014AJ....148..118N} for their sample and 
is further enhanced by the inclusion of
the sample of objects from \citet{2012MNRAS.421.2616N}.  The
procedures noted above should have corrected our estimated true mass
distribution for this bias.

In the bottom panel of \autoref{m1dist} there is good agreement
between the estimated true current red giant mass distribution and the
equivalent simulated mass distribution. Note that the simulation only
considers stars with initial masses up to 3\msolar~so we can only
compare the observed and simulated mass distributions up to this mass.
A characteristic property of both distributions is that each shows a
peak at a current red giant mass of $\sim$1.5\msolar.  This provides
direct confirmation that the mass distribution of the primary masses in binary
systems with periods of $\sim$100--1000 days is similar to that of the 
single stars in the LMC.  The single star mass distribution in the LMC 
is dominated but a burst of star formation in the interval 0.5--4 Gyr ago 
and this star formation history was used in the models of
\citet{2012MNRAS.423.2764N} as well as here.

\subsection{The mass ratio distribution} 

\begin{figure}
\begin{center}
\includegraphics[width={1.0\columnwidth}]{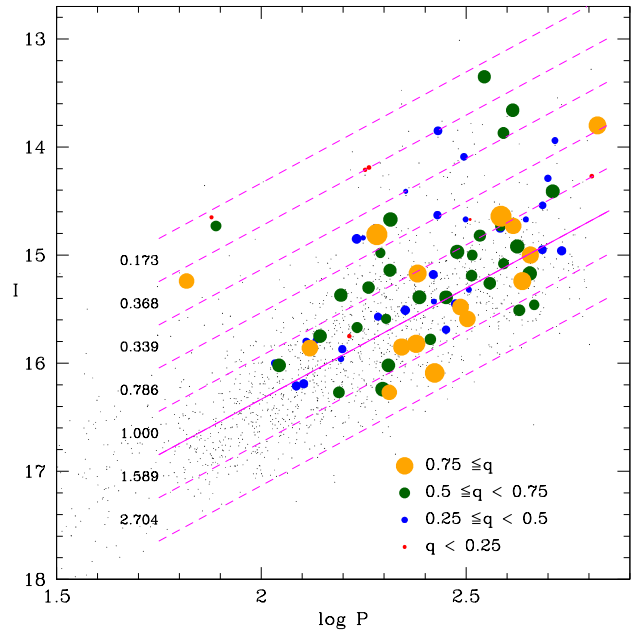}
\caption{The same as \autoref{i-logp+m1} except that the colored symbols indicate
the $q$ value.}
\label{i-logp+q}
\end{center}
\end{figure}

\begin{figure}
\begin{center}
\includegraphics[width={1.0\columnwidth}]{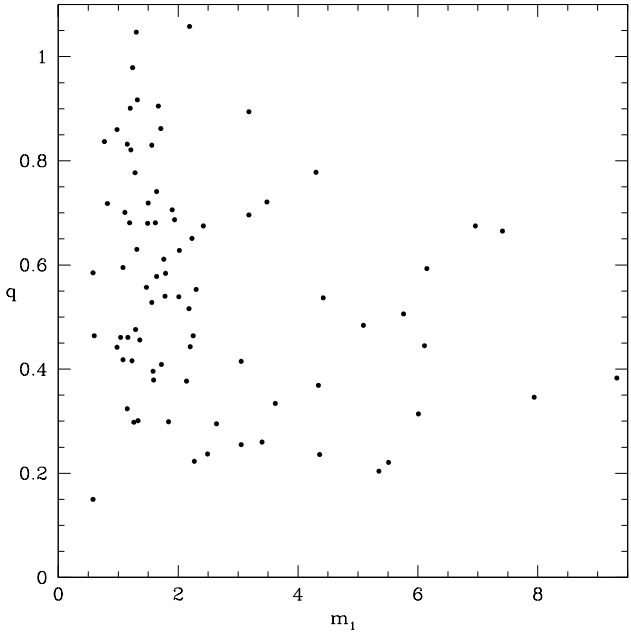}
\caption{The $q$ value plotted against $m_{\rm 1}$ for the sequence E stars in
our sample.}
\label{q-m1}
\end{center}
\end{figure}

In \autoref{i-logp+q} we show the mass ratio $q = m_{\rm 2}/m_{\rm 1}$
for our sample plotted on the $I$-$\log P$ plane in order to see if
there is any preferential position for large or small $q$ values that
could lead to a selection bias, as was found in the case of $m_{\rm
  1}$.  The different $q$ values appear to be spread 
uniformly across the $I$-$\log P$ sequence formed by the complete OGLE
II sample of \citet{2004AcA....54..347S}, unlike the masses $m_{\rm
  1}$.  In \autoref{q-m1}, $q$ is plotted against $m_{\rm 1}$
to show the dependence of $q$ on $m_{\rm 1}$.  
A two-sample KS test on the data
divided into two at $m_{\rm 1} < 3$\msolar~and $m_{\rm 1} \ge
3$\msolar~ gives a probability of 10\% that the more and less massive
samples come from the same underlying distribution, which implies 
that there is a reasonable probability that binaries with $m_{\rm 1} \ge                                                                                                                                           
3$\msolar~have lower $q$ values than binaries with less massive primaries.  Similarly,
comparing the systems where the primary star was originally an F or G
star ($m_{\rm 1} < 1.4$\msolar) with the systems where the primary was 
originally an A or late-B star ($3 > m_{\rm 1} > 1.4$\msolar) 
gives a probability of 47\% that
the two samples come from the same underlying distribution.  Thus we
have no evidence that $q$ depends on $m_{\rm 1}$ in the mass range
$m_{\rm 1} < 3$\msolar.

\begin{figure}
\begin{center}
        \includegraphics[width={1.0\columnwidth}]{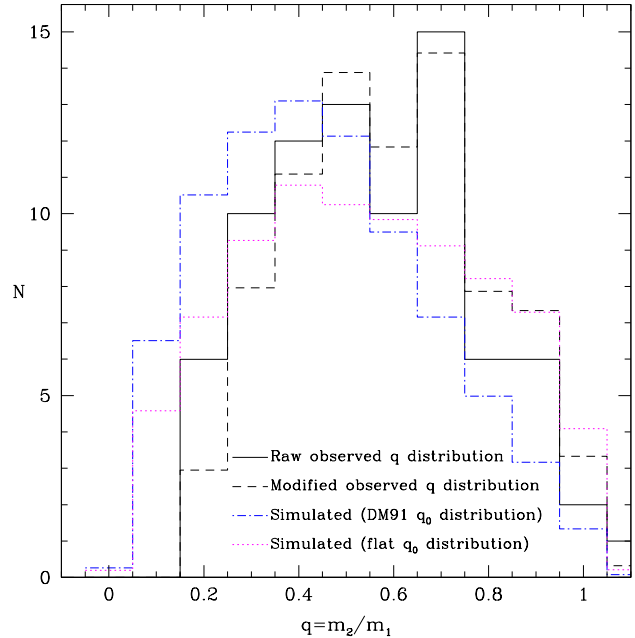}
\caption{Distribution of the mass ratio $q$. The solid black line shows the raw
distribution for our sample while the dashed black line
shows the distribution adjusted for possible selection bias as described in the text.
Also shown is the predicted $q$ distribution for 
according to a simulation based on the method of \citet{2012MNRAS.423.2764N}
(blue dot-dash line) and a similar simulation except that a flat initial $q$ 
distribution was used (pink dotted line). All distributions 
are normalized to have the same total number of 
counts as the raw observed distribution.
\label{qdist}}
\end{center}
\end{figure}

The raw distribution of $q$ values for our complete sample is shown in
\autoref{qdist}.  In an attempt to remove any bias due to the selection 
of bright objects, we applied the same weighting scheme to the objects 
that was used for the $m_{\rm 1}$ values in order to get a distribution 
that is selected uniformly from the sample of
\citet{2004AcA....54..347S} in the $I$-$\log P$ plane in the interval 
$1.75 < \log(P) = 2.85$.  This
distribution is also shown in \autoref{qdist}.

We aim to compare the $q$ distribution for our sample of red giant
binaries with other observed distributions.  The two most comprehensive
observed binary mass ratio distributions are those of
\citet{1991A&A...248..485D} and \citet{2010ApJS..190....1R} which
apply to stars in the solar vicinity.  In both these studies, the
primary stars are on the main sequence whereas in our case the primary
star is a red giant that has undergone some mass loss so that the
current mass ratio is different from that on the main sequence.  Also,
the detection of ellipsoidal variability, and hence the lifetime of a
sequence E star, depends on the mass ratio.  These factors need to
be taken into account when comparing our sample to those of
\citet{1991A&A...248..485D} and \citet{2010ApJS..190....1R}.  We do a
comparison with the \citet{1991A&A...248..485D} distribution by
running a simulation based on the prescriptions of
\citet{2012MNRAS.423.2764N} as described in \autoref{s:m1-dist}.  This
simulation estimates the distribution of $q$ values of sequence E
stars in the full sample of \citet{2004AcA....54..347S} when starting
on the main sequence with the $q$ distribution of
\citet{1991A&A...248..485D}.  We used the fit to the $q$ distribution
given by \citet{1991A&A...248..485D} which is a Gaussian curve with a
mean of 0.23 and a dispersion of 0.42 (their Figure 10).  A similar
simulation was made using a flat main sequence distribution for $q$,
which \citet{2010ApJS..190....1R} suggest as a good approximation to
their observed $q$ distribution for $0.2 < q < 0.95$.  The
distributions resulting from these two simulations are shown in
\autoref{qdist} along with our observed distribution of $q$.

Our raw distribution of $q$ values differs significantly from each of
the simulations.  A KS test gives a probability of less than 
1.6$\times$10$^{-6}$ that our raw distribution and the simulated distribution
using the main sequence $q$ distribution of
\citet{1991A&A...248..485D} come from the same underlying distribution.
Similarly, we find a probability of less than 0.006 that our raw
distribution and the simulated distribution using the main sequence $q$
distribution of \citet{2010ApJS..190....1R} come from the same
underlying distribution.  

Looking at \autoref{qdist}, it is clear that our raw and bias-corrected distributions
are shifted to higher $q$ values than predicted by the models which use 
the input distribution of \citet{1991A&A...248..485D}.  Our distributions
are better aligned with that resulting from use of
the distribution of \citet{2010ApJS..190....1R}.  We do
not find any evidence for an excess of similar-mass binaries with $q >
0.95$, in agreement with the findings of \citet{1991A&A...248..485D}
but not with \citet{2010ApJS..190....1R} who do find such an excess.

\subsection{The eccentricities}

\begin{figure}
\begin{center}
\includegraphics[width={1.0\columnwidth}]{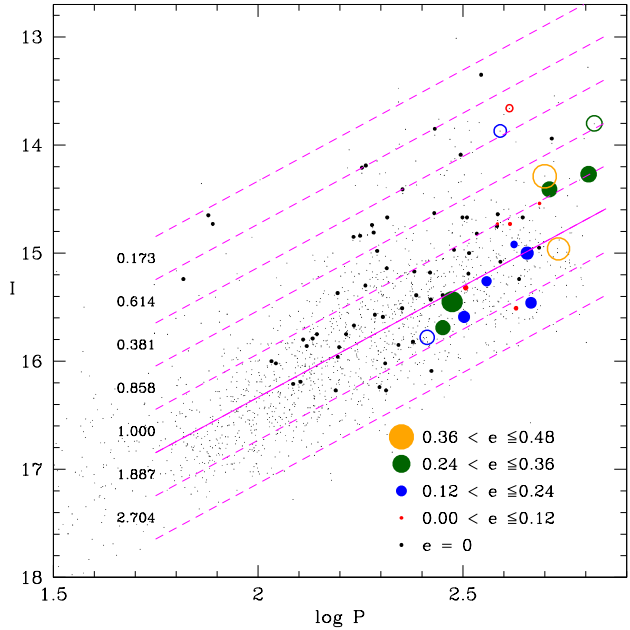}
\caption{The same as \autoref{i-logp+m1} except that the colored symbols indicate
the value of the eccentricity $e$.  Open symbols belong to the seven objects studied by 
\citet{2012MNRAS.421.2616N}.}
\label{i-logp+e}
\end{center}
\end{figure}

\begin{figure}
\begin{center}
\includegraphics[width={1.0\columnwidth}]{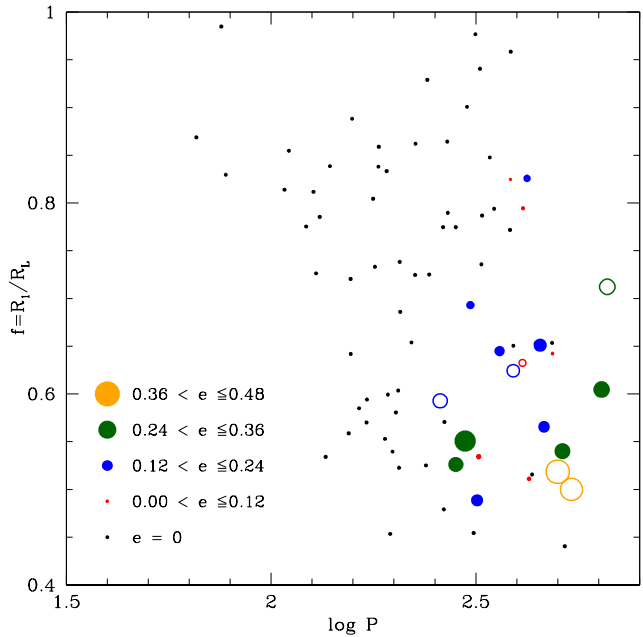}
\caption{Our observed sample of ellipsoidal variables in the $f$-$\log P$ plane.
The symbols are the same as in \autoref{i-logp+e}.
}
\label{frl-logp+e}
\end{center}
\end{figure}

In the study of the OGLE II ellipsoidal variables by
\citet{2004AcA....54..347S} it was noted that about 10\% of 
ellipsoidal variables had non-sinusoidal light curves.  
\citet{2004AcA....54..347S} used simple binary light curve models 
to show that similar light curves could be caused by ellipsoidal 
variables in eccentric orbits. However, \citet{2004AcA....54..347S} 
were not able to derive eccentricities as they did not have any radial 
velocity data and they relied on the light curves alone when deciding 
if an object had an eccentric orbit. The modelling of both the light and radial 
velocity curves done by \citet{2012MNRAS.421.2616N}  confirmed that 
non-sinusoidal light curve shapes of ellipsoidal variables are indeed caused 
by eccentric orbits.

If we exclude from our sample the seven objects studied by \citet{2012MNRAS.421.2616N} 
which were chosen because they had light curves suggesting eccentric 
orbits, we find that 15 out of 
74, or $\sim$20\% of the systems have eccentric orbits, a higher fraction than 
estimated by \citet{2004AcA....54..347S}.  Examination of the list of objects 
with eccentric orbits given by \citet{2004AcA....54..347S} shows that only
8 of our 16 eccentric objects were considered eccentric by
\citet{2004AcA....54..347S}.  Light and velocity curve modelling, as
we have done, clearly reveals more eccentric objects, so the true
fraction of eccentric orbits is likely to be roughly 20\%, or twice
the fraction estimated by \citet{2004AcA....54..347S}.

The eccentricities of the objects in our sample are indicated as a function of
position on the $I$-$\log P$ plane in \autoref{i-logp+e}.  Because the objects
studied by \citet{2012MNRAS.421.2616N} were deliberately selected to be
eccentric, they are shown as open symbols.  
The objects in \autoref{i-logp+e} which have eccentric orbits
(excluding the eccentricity-selected objects of
\citealt{2012MNRAS.421.2616N}) tend to lie on the fainter, longer
period side of the mean $I$--$\log P$ relation for ellipsoidal red
giant variables.  This result can also be seen in Figure 5 of
\citet{2004AcA....54..347S}.  This means that our estimate of 20\% for
the fraction of ellipsoidal red giant variables that are eccentric may
be a lower limit since our sample of objects is biased towards the
objects lying above the mean relation in the $I$-$\log P$ plane.
If we correct for the observational bias to brighter objects using
the method applied above for the distributions of $m_1$ and $q$,
but using the restricted sample containing objects from \citet{2014AJ....148..118N}
and \cite{2010MNRAS.405.1770N} only, we estimate that 28\%
of ellipsoidal variables with $1.75 < \log(P) < 2.85$ should have
eccentric orbits.

The red giant ellipsoidal variables are substantially filling their
Roche lobes and these binaries will be subject to tidal forces tending
to circularize their orbits.  \autoref{i-logp+e} and Figure 5 of
\citet{2004AcA....54..347S} show that these ellipsoidal variables only
have non-zero eccentricity for orbital periods $P \ga 200$\,days.
This is similar to the situation for barium stars
\citep{1998A&A...332..877J} which are binaries that in the past are
thought to have substantially filled their Roche lobes and undergone
mass transfer by an enhanced stellar wind
\citep[e.g.][]{2000MNRAS.316..689K}.  However, spectroscopic binaries
containing a red giant, but where the Roche lobe is not necessarily
substantially filled, show eccentric orbits down to orbital periods of
$\sim$10\,days \citep{1993A&A...271..125B}.  
Those systems presumably lie below the fainter edge of the
$I$--$\log P$ sequence for ellipsoidal variables in \autoref{i-logp+e}.  Here, 
their very low amplitude ellipsoidal variability is not detectable by
the OGLE observations.

As noted above, the objects which have eccentric orbits
tend to lie on the fainter, longer period
side of the mean $I$--$\log P$ relation for ellipsoidal red giant
variables.  The longer periods of these objects at a
given $I$ means that they will tend to have larger orbital separations
and hence smaller values of $R_{\rm 1}/a$ than their shorter period counterparts
with circular orbits.  Since the tidal
circularization time varies as $(R_{\rm 1}/a)^{-8}$
\citep[e.g.][]{2000A&A...357..557S}, this means that the systems with
longer periods should have longer circularization times and are hence
they are more likely to have eccentric orbits, as observed.  Note that
the Roche lobe filling factor is $f = R_{\rm 1}/R_{\rm L} = h(q) R_{\rm
  1}/a$ (where $h(q)$ can be obtained from \autoref{egg_eqn} above), 
so that binaries with smaller values of $f$ will have
smaller values of $(R_{\rm 1}/a)$ and hence should have longer
circularization times and be more likely to have eccentric orbits.
This is seen to be the case in \autoref{frl-logp+e} which shows that
in the $f$-$\log P$ plane, the detectable ellipsoidal binary systems
with eccentric orbits tend to have small $f$ values, as well as long
orbital periods.

\section{Implications for tidal interaction theories} \label{tidal}

\begin{figure}
\begin{center}
\includegraphics[width={1.0\columnwidth}]{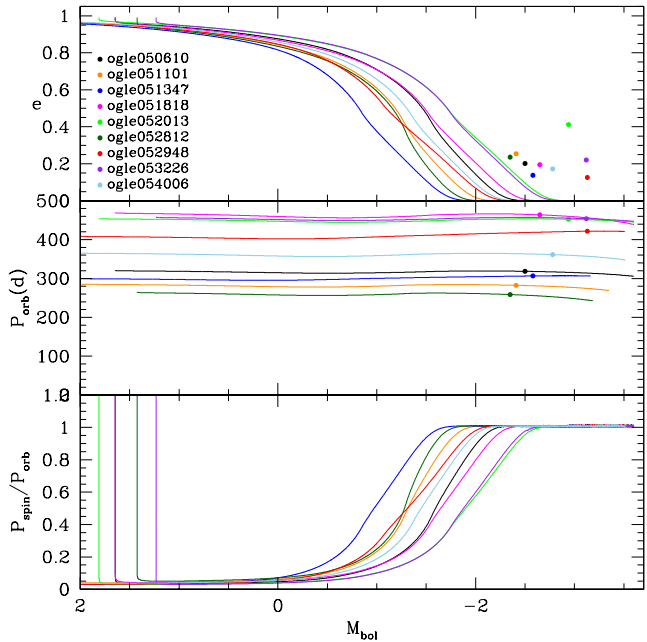}
\caption{
The variation of eccentricity, orbital period and the ratio of spin to
orbital period for low mass ($m_1<$1.85\msolar) stars as they evolve up the RGB.
Current values of eccentricity and orbital period are shown by dots of different
color for nine of our observed systems (we have no rotational velocity measurements
so there are no observational points in the bottom panel).  The lines show simulations for each
of the nine systems, with line color matching the dot color.  See text for details.
}
\label{e.evolution}
\end{center}
\end{figure}

We have derived complete orbital solutions for 81 binary systems
containing a red giant primary with a luminosity in the upper two
magnitudes of the RGB, and 22 of these systems currently have a
non-zero eccentricity.  Since we know all the properties of this
unique set of 22 eccentric binary systems, we can use them to test
tidal theories by seeing if these theories allow the observed non-zero
eccentricity to exist.  To do this, we selected the 9 systems 
(seven from \citealt{2014AJ....148..118N} and two from \citealt{2012MNRAS.421.2616N}) in our
sample that have current eccentricities $e > 0.1$ and red giant
primary masses $m_1<$ 1.85\msolar~so that they are evolving up the low
mass RGB.  The evolution of these systems was followed using the
prescriptions in the standard model of \citet{2012MNRAS.423.2764N},
together with the assumption of non-zero eccentricity.  The effects of
tides on eccentricity, orbital period and spin period of the red giant
are taken into account with the widely-used formulae of
\citet{1977A&A....57..383Z,1989A&A...220..112Z} and
\citet{1981A&A....99..126H}, as implemented in
\citet{2002MNRAS.329..897H}.  Specifically, we use Equations 6, 11,
20, 25, 26 and 30--35 from the latter paper.  The calculations were
started at the base of the RGB, where it was assumed that the red
giant had no spin angular momentum and that the eccentricity was very
large (a value of 0.99 was used).  This latter assumption was made in
order to produce the maximum possible value of $e$ at all stages of
evolution up the RGB.  The initial orbital period and masses were
adjusted so that the system had its observed period and derived masses
at its current observed red giant luminosity.

The results of these evolution calculations for each of the 9 modelled
systems are shown in \autoref{e.evolution}.  The base of the giant
branch was fainter than the plotted luminosity range except for the four
most massive stars which have masses close to 1.8\msolar~and
luminosities $M_{\rm bol} \approx 1$ at the base of the giant branch.
It can be seen that at low luminosities ($M_{\rm bol} \ga 0$) the
eccentricity is large and has not changed much from the initial value
while the spin period is much shorter than the orbital period.  This
short spin period is a result of the fact that there is strong
spin-orbit coupling which has spun up the red giant very quickly (see
the rapid decrease in the spin period in the bottom panel of
\autoref{e.evolution} for the 3 most massive stars).  This rapid spin
up, and the small ratio of $P_{\rm spin}/P_{\rm orb}$, is due to the
high eccentricity which means that the periastron distance is small, and
at periastron tidal effects are large and the orbital velocity is high
thereby causing the rapid spin up of the red giant.  In all
cases, as the red giant starts to substantially fill its Roche lobe
between $M_{\rm bol} = -0.5$ and $-2.5$, tides effectively circularize
the orbit and the spin and orbital periods equalize.  As a result of
the latter effect, some spin angular momentum is fed back into the
orbit and a small increase in orbital period can be seen.

\begin{figure}[t]
\begin{center}
\includegraphics[width={1.0\columnwidth}]{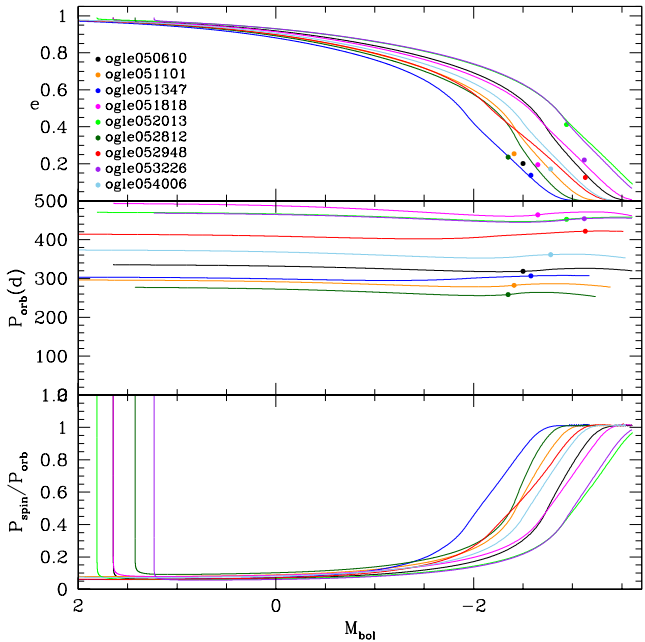}
\caption{
The same as \autoref{e.evolution} but using a tidal damping factor $F_{\rm tid} = 0.01$.
}
\label{e.evolution_ftide0.01}
\end{center}
\end{figure}

The most obvious outcome of these calculations is that, in every case,
the models predict that the observed systems should have been
circularized ($e = 0$) by the time they reached the luminosities they
are observed to have today, yet the observed eccentricities are
clearly non-zero.  Given that the models were started with a very high
eccentricity, this suggests that the tidal theories we have used for
convective red giant stars greatly overestimate the rate of tidal
circularization.  Although mechanisms have been proposed that can
maintain or boost eccentricity, such as the external disk mechanism
\citep{1991ApJ...370L..35A,1996A&A...314L..17W} or mass transfer
preferentially at periastron \citep{2000A&A...357..557S}, these
mechansims would require a substantial amount of circumbinary dust
whereas mid-infrared photometry suggests that such circumbinary dust
does not exist in these systems \citep{2010MNRAS.405.1770N}.  Note
that our finding that tidal circularization rates are {\em
  overestimated} in red giants is the reverse of the finding for main
sequence binaries where it is estimated that the tidal circularization
rates are {\em underestimated} by factors of 50 to 100
\citep{2005ApJ...620..970M, 2008ApJS..174..223B,2013AJ....145....8G}.
To test how much the tidal circularization rates for binaries with a red
giant primary would need to be adjusted to get agreement with observed
eccentricities, we follow \citet{2008ApJS..174..223B} and introduce a
factor $F_{\rm tid}$ which multiplies the rates of change of
eccentricity and spin due to tides (Equations 25 and 26 of
\citealt{2002MNRAS.329..897H}) and re-run the simulation which was shown
in \autoref{e.evolution}.  In \autoref{e.evolution_ftide0.01}, we show
the results for $F_{\rm tid} = 0.01$.  This value of $F_{\rm tid}$
producess reasonable agreement between the observed and predicted
eccentricities, at least when assuming that the initial eccentricity
is large.  Our estimated value of $F_{\rm tid}$ is a factor of
$\sim$$10^4$ smaller than the values of 50--100 estimated for main
sequence stars by \cite{2008ApJS..174..223B} and
\cite{2013AJ....145....8G}.  It seems that severe modifications of
tidal theories in convective stars are required in order to be able to
model stars with widely different structures using a single theory.

\section[]{Summary and Conclusions}

We have carried out orbital modelling for 81 ellipsoidal red giant
binaries with the 2010 version of the WD code.  The working sample was
collected from the works of \citet{2014AJ....148..118N},
\citet{2010MNRAS.405.1770N} and \citet{2012MNRAS.421.2616N} and it
contains 59 systems with circular orbits and 22 systems with eccentric
orbits.  Using this sample, and correcting for
selection bias, we compare the derived distributions of primary
mass $m_{\rm1}$ and mass ratio $q = m_{\rm2}/m_{\rm 1}$ with model
predictions for these distributions based on the methods in
\citet{2012MNRAS.423.2764N}.  We find values of $m_{\rm1}$ from
0.6--9\msolar ~and a peak in the distribution of $m_{\rm1}$ near
1.5\msolar, consistent with studies of the LMC star formation history
which show a burst of star formation starting $\sim$4 Gyr ago.  The
derived distribution of $q$ is in better agreement with
the flat $q$ distribution derived for the solar vicinity by \citet{2010ApJS..190....1R}
than it is with the solar vicinity $q$ distribution derived by \citet{1991A&A...248..485D}
which favors lower $q$ values.  We found that about 20\%  of red
giant ellipsoidal binaries have eccentric orbits, twice the
fraction found by \citet{2004AcA....54..347S} in their OGLE II sample.
We also note that half the objects found to have eccentric orbits
after modelling the light and velocity curves were not noted by
\citet{2004AcA....54..347S} to be eccentric as a result of examination
of the light curve alone.  The complete parameter sets obtained for
the eccentric red giant binary systems were used to test standard
theories of tidal interaction in eccentric binaries containing a
convective star \citep{1977A&A....57..383Z,
  1989A&A...220..112Z,1981A&A....99..126H}.  It was found that the
tidal circularization rates predicted by these theories are about
100 times {\em faster} than allowed by the existence of
eccentric red giant binaries on the upper parts of the RGB.  This
contrasts with studies of convective main sequence binaries where the
tidal circularization rates predicted by the standard
theories are about 50--100 times {\em slower} than observations suggest.
We conclude that current theories of tidal dissipation in convective
stars need to be substantially improved.

\section*{Acknowledgements}
JDN is supported by the National Natural Science Foundation of China (NSFC) through grant 11303043 and the 
Young Researcher Grant of National Astronomical Observatories, Chinese Academy of Sciences.
PRW was partially supported in this work by the Australian Research Council's Discovery Projects funding scheme 
(project number DP120103337).  CPN is a Lise Meitner fellow and acknowledges support by the Austrian
Science Fund (FWF) under project number M 1696-N27.  The authors would like to thank Stephen Justham for 
useful discussions regarding observational tests of tidal theories.  We would also like to
thank the referee for helping us improve the presentation of the paper.


\begin{thebibliography}{80}
\bibitem[Alencar \& Vaz(1997)]{1997A&A...326..257A} Alencar, S.~H.~P., \& Vaz, L.~P.~R.\ 1997, \aap, 326, 257 
\bibitem[Artymowicz et al.(1991)]{1991ApJ...370L..35A} Artymowicz, P., Clarke, C.~J., Lubow, S.~H., \& Pringle, J.~E.\ 1991, \apjl, 370, L35 
\bibitem[Belczynski et al.(2008)]{2008ApJS..174..223B} Belczynski, K., Kalogera, V., Rasio, F.~A., et al.\ 2008, \apjs, 174, 223 
\bibitem[Bertelli et al.(2008)]{2008A&A...484..815B} Bertelli, G., Girardi, L., Marigo, P., \& Nasi, E.\ 2008, \aap, 484, 815 
\bibitem[Boffin et al.(1993)]{1993A&A...271..125B} Boffin, H.~M.~J., Cerf, N., \& Paulus, G.\ 1993, \aap, 271, 125
\bibitem[Cutri et al.(2003)]{2003yCat.2246....0C} Cutri, R.~M., Skrutskie, M.~F., van Dyk, S., et al.\ 2003, VizieR Online Data Catalog, 2246, 0 
\bibitem[de Grijs et al.(2014)]{2014AJ....147..122D} de Grijs, R., Wicker, J.~E., \& Bono, G.\ 2014, \aj, 147, 122
\bibitem[Dopita et al. (2007)]{2007Ap&SS.310..255D} Dopita M., Hart J., McGregor P., Oates P., Bloxham G., Jones D., 2007, Ap\&SS, 310, 255 
\bibitem[Dopita et al. (2010)]{2010Ap&SS.327..245D} Dopita M., et al., 2010, Ap\&SS, 327, 245 
\bibitem[Duquennoy \& Mayor(1991)]{1991A&A...248..485D} Duquennoy A., Mayor M., 1991, APP, 248, 485
\bibitem[Eggleton (1983)]{1983ApJ...268..368E} Eggleton P.~P., 1983, ApJ, 268, 368
\bibitem[Fraser et al.(2008)]{2008AJ....136.1242F} Fraser, O.~J., Hawley, S.~L., \& Cook, K.~H.\ 2008, \aj, 136, 1242 
\bibitem[Geller et al.(2013)]{2013AJ....145....8G} Geller, A.~M., Hurley, J.~R., \& Mathieu, R.~D.\ 2013, \aj, 145, 8 
\bibitem[Han et al.(2002)]{2002ASPC..279..297H} Han, Z., Podsiadlowski, P., \& Tout, C.~A.\ 2002, Exotic Stars as Challenges to Evolution, 279, 297 
\bibitem[Haario et al.(2001)]{haario2001} Haario, H., Saksman, E. \& Tamminen, J.\ 2001, \textit{Bernoulli}, 7(2), 223
\bibitem[Haario et al.(2006)]{haario2006} Haario, H., Laine, M., Mira, A. \& Saksman, E.\ 2006, Stat Comput, 16, 339
\bibitem[Houdashelt et al.(2000a)]{2000AJ....119.1424H} Houdashelt M.~L., Bell R.~A., Sweigart A.~V., Wing R.~F., 2000a, \aj, 119, 1424
\bibitem[Houdashelt et al.(2000b)]{2000AJ....119.1448H} Houdashelt M.~L., Bell R.~A., Sweigart A.~V., 2000b, \aj, 119, 1448
\bibitem[Hurley et al.(2002)]{2002MNRAS.329..897H} Hurley, J.~R., Tout, C.~A., \& Pols, O.~R.\ 2002, \mnras, 329, 897 
\bibitem[Hut(1981)]{1981A&A....99..126H} Hut, P.\ 1981, \aap, 99, 126 
\bibitem[Ita et al.(2004)]{2004MNRAS.353..705I} Ita, Y., Tanab{\'e}, T., Matsunaga, N., et al.\ 2004, \mnras, 353, 705
\bibitem[Jorissen et al.(1998)]{1998A&A...332..877J} Jorissen, A., Van Eck, S., Mayor, M., \& Udry, S.\ 1998, \aap, 332, 877
\bibitem[Kallinger et al.(2010)]{2010A&A...522A...1K} Kallinger, T., Mosser, B., Hekker, S., et al.\ 2010, \aap, 522, A1
\bibitem[Karakas et al.(2000)]{2000MNRAS.316..689K} Karakas, A.~I., Tout, C.~A., \& Lattanzio, J.~C.\ 2000, \mnras, 316, 689
\bibitem[Keller \& Wood(2006)]{2006ApJ...642..834K} Keller, S.~C., \& Wood, P.~R.\ 2006, \apj, 642, 834 
\bibitem[Latham et al.(2002)]{2002AJ....124.1144L} Latham, D.~W., Stefanik, R.~P., Torres, G., et al.\ 2002, \aj, 124, 1144 
\bibitem[Madore (1982)]{1982ApJ...253..575M} Madore B.~F., 1982, ApJ, 253, 575 
\bibitem[Meibom \& Mathieu(2005)]{2005ApJ...620..970M} Meibom, S., \& Mathieu, R.~D.\ 2005, \apj, 620, 970
\bibitem[Morris(1985)]{1985ApJ...295..143M} Morris, S.~L.\ 1985, \apj, 295, 143 
\bibitem[Nicholls et al.(2010)]{2010MNRAS.405.1770N} Nicholls, C.~P., Wood, P.~R., \& Cioni, M.-R.~L.\ 2010, \mnras, 405, 1770 
\bibitem[Nicholls \& Wood(2012)]{2012MNRAS.421.2616N} Nicholls, C.~P., \& Wood, P.~R.\ 2012, \mnras, 421, 2616 
\bibitem[Nie et al.(2012)]{2012MNRAS.423.2764N} Nie, J.~D., Wood, P.~R., \& Nicholls, C.~P.\ 2012, \mnras, 423, 2764 
\bibitem[Nie \& Wood(2014)]{2014AJ....148..118N} Nie, J.~D., \& Wood, P.~R.\ 2014, \aj, 148, 118 
\bibitem[Pawlak et al.(2014)]{2014AcA....64..293P} Pawlak, M., Soszy{\'n}ski, I., Pietrukowicz, P., et al.\ 2014, \actaa, 64, 293 
\bibitem[Piatti \& Geisler(2013)]{2013AJ....145...17P} Piatti, A.~E., \& Geisler, D.\ 2013, \aj, 145, 17
\bibitem[Raghavan et al.(2010)]{2010ApJS..190....1R} Raghavan, D., McAlister, H.~A., Henry, T.~J., et al.\ 2010, \apjs, 190, 1 
\bibitem[Soker(2000)]{2000A&A...357..557S} Soker, N.\ 2000, \aap, 357, 557
\bibitem[Soszynski et al.(2004)]{2004AcA....54..347S} Soszynski, I., Udalski, A., Kubiak, M., et al.\ 2004, \actaa, 54, 347 
\bibitem[Soszynski et al.(2007)]{2007AcA....57..201S} Soszynski, I., Dziembowski, W.~A., Udalski, A., et al.\ 2007, \actaa, 57, 201 
\bibitem[Stello et al.(2013)]{2013ApJ...765L..41S} Stello, D., Huber, D., Bedding, T.~R., et al.\ 2013, \apjl, 765, L41 
\bibitem[Szymanski(2005)]{2005AcA....55...43S} Szymanski, M.~K.\ 2005, \actaa, 55, 43
\bibitem[Udalski et al.(1997)]{1997AcA....47..319U} Udalski, A., Kubiak, M., \& Szymanski, M.\ 1997, \actaa, 47, 319 
\bibitem[Waelkens et al.(1996)]{1996A&A...314L..17W} Waelkens, C., Van Winckel, H., Waters, L.~B.~F.~M., \& Bakker, E.~J.\ 1996, \aap, 314, L17
\bibitem[Wilson (1979)]{1979ApJ...234.1054W} Wilson R.~E., 1979, ApJ, 234, 1054 
\bibitem[Wilson (1990)]{1990ApJ...356..613W} Wilson R.~E., 1990, ApJ, 356, 613 
\bibitem[Wilson \& Devinney(1971)]{1971ApJ...166..605W} Wilson, R.~E., \& Devinney, E.~J.\ 1971, \apj, 166, 605
\bibitem[Wilson et al.(2009)]{2009ApJ...702..403W} Wilson, R.~E., Chochol, D., Kom{\v z}{\'{\i}}k, R., et al.\ 2009, \apj, 702, 403 
\bibitem[Wood et al.(1999)]{1999IAUS..191..151W} Wood, P.~R., Alcock, C., Allsman, R.~A., et al.\ 1999, Asymptotic Giant Branch Stars, IAUS, 191, 151
\bibitem[Wood(2000)]{2000PASA...17...18W} Wood, P.~R.\ 2000, \pasa, 17, 18 
\bibitem[Wood(2015)]{2015MNRAS.448.3829W} Wood, P.~R.\ 2015, \mnras, 448, 3829 
\bibitem[Zahn(1977)]{1977A&A....57..383Z} Zahn, J.-P.\ 1977, \aap, 57, 383
\bibitem[Zahn(1989)]{1989A&A...220..112Z} Zahn, J.-P.\ 1989, \aap, 220, 112
\end{thebibliography}
\end{document}